\documentclass[a4paper,12pt]{article}
\usepackage{amsmath}
\usepackage{graphicx}
\usepackage{amsfonts}

\DeclareFontFamily{U}{rsf}{}
\DeclareFontShape{U}{rsf}{m}{n}{
  <5> <6> rsfs5 <7> <8> <9> rsfs7 <10-> rsfs10}{}
\DeclareMathAlphabet\Scr{U}{rsf}{m}{n}

\mathchardef\varGamma="0100
\mathchardef\varDelta="0101
\mathchardef\varTheta="0102
\mathchardef\varLambda="0103
\mathchardef\varXi="0104
\mathchardef\varPi="0105
\mathchardef\varSigma="0106
\mathchardef\varUpsilon="0107
\mathchardef\varPhi="0108
\mathchardef\varPsi="0109
\mathchardef\varOmega="010A

\title{ Stringy Instanton Effects in Magnetised D-brane Models}
\begin{document}

\thispagestyle{empty}

\vspace{-2cm}

\begin{flushright}
{\small
CERN-TH-PH/2009-023\\
CPHT-RR010.0209\\
DFTT 35/2009\\
LPT-ORSAY 09-09
}
\end{flushright}
\vspace{1cm}

\begin{center}
{\bf\Large Stringy Instanton Effects \\[18pt]
 in Models with Rigid Magnetised D-branes}

\vspace{0.8cm}

{\bf
Carlo Angelantonj$^{*,}\footnote{On leave of absence from Department of Theoretical Physics, University of Turin, and INFN Sezione di Torino, Italy}$,
Cezar Condeescu$^{+,}\footnote{On leave of absence from Institute of
Mathematics 'Simion Stoilow' of the Romanian
Academy,\\
P.O. Box 1-764, Bucharest, RO-70700, Romania.}$,
Emilian Dudas$^{+,\dag}$ \\[8pt]
and Michael Lennek$^{+}$  \vspace{0.8cm}
}

{\it $^*$ Theory Group CERN, 1211 Geneva 23, Switzerland\\[12pt]
 $^{+}$ Centre de Physique Th\'eorique,
Ecole Polytechnique and CNRS,\\
F-91128 Palaiseau, France.\\[12pt]
$^{\dag}$ LPT,Bat. 210, Univ. de Paris-Sud, F-91405 Orsay, France.}

\vspace{0.7cm}

{\bf Abstract}
\end{center}
\vspace{.2cm}
We compute instantonic effects in globally consistent $T^6/\mathbb{Z}_2\times \mathbb{Z}_2$ orientifold models
with discrete torsion and magnetised D-branes. We consider fractional
branes and instantons wrapping the same rigid cycles.  We clarify and analyse
in detail the low-energy effective action on D-branes in these models.
We provide explicit examples where instantons induce linear terms in the charged fields,
or non-perturbative mass terms are generated. We also find examples where the gauge theory on fractional branes has
conformal symmetry at one-loop, broken by instantonic mass terms at a
hierarchically small energy scale.

\clearpage

\begin{flushright}
{\today}
\end{flushright}

\tableofcontents

\section{Introduction}

String compactifications with internal magnetic fields and/or branes at angles were proposed some
time ago as a different way to (partially) break spacetime supersymmetry \cite{magnetic}. Explicit CFT constructions \cite{intersecting} provided a simple and geometrically intuitive framework for constructing vacuum configurations as close as possible to the Standard Model, and its supersymmetric extension.  For some applications of such models to moduli stabilization see \cite{stabmod}.

Although, most of the efforts in intersecting brane constructions were devoted to the simple four-dimensional
compactification on $T^6 /\mathbb {Z}_2 \times \mathbb {Z}_2$ orbifolds, many unwanted features plagued the original constructions.
For instance, the massless open-string spectrum on this orbifold
always contained three chiral multiplets in the
adjoint representations of the gauge group. Their presence is actually
related to the possibility of displacing the
D-branes at arbitrary positions along the internal manifold or, more
geometrically, to the fact that the intersecting
branes wrap non-rigid cycles of the $T^6 /\mathbb {Z}_2 \times \mathbb
{Z}_2$. Although, it has been shown in
\cite{SSmagnetic} how a proper use of Scherk-Schwarz deformations
might make these adjoint fermions massive
(at the price, however, of breaking supersymmetry), a general
mechanism for lifting these states compatibly
with space-time supersymmetry is still missing.

$\mathbb{Z}_N \times \mathbb {Z}_M$ orbifold compactifications come,
however, in inequivalent versions, due to the possibility of turning
on non-vanishing, localised, background values for the antisymmetric
NS-NS tensor field. In turn, this corresponds
to introducing a non-trivial phase (a minus sign for the case $N=M=2$)
for disconnected orbits in the twisted sector
\cite{discretetorsion}, thus changing the Hodge structure of the
compact manifold. If an orientifold projection is combined with
discrete torsion, exotic O-planes with positive tension and charge are
present, at least for the $\mathbb {Z}_2 \times \mathbb {Z}_2$ case
\cite{aadds}. In the most naive set-up, the RR tadpole cancellation
for these O-planes calls for the introduction of anti-brane, therefore
providing a realisation of Brane Supersymmetry Breaking (BSB)
\cite{BSB}. However, in \cite{cristina,bcms} it has been
shown that fully supersymmetric orientifold vacua actually do exist
that involve (intersecting) fractional branes wrapping so-called rigid cycles
\footnote{Supersymmetric models with bulk D-branes were previously constructed
in \cite{ms}.}. In these constructions, chiral multiplets in adjoint
representations are absent because of the rigid nature
of the cycles, that therefore cannot be continuously deformed. In
other words, the combination of the orbifold and orientifold
twists has a non-trivial action and projects away the internal Wilson
lines or, in the T-dual picture, the coordinates of
the D-branes in the compact space. Early examples of orientifold models with branes intersecting rigid cycles were built in
\cite{aadds, larosa}.

Although intersecting brane models are very useful tools to build
chiral lower-dimensional vacua with unitary gauge groups while also
 providing a geometrical origin for the family replication observed for
  charged matter, it has become clear that they
fail to capture most phenomenological properties of MSSM. This
limitation, however, mainly resides in the perturbative
approach since, starting from the pioneering paper on instantonic
contributions in string theory \cite{inst}, it has been
shown that non-perturbative effects cannot sometimes be neglected and
induce interesting features in orientifold vacua.
For instance, they provide a way of generating (otherwise forbidden) new terms in the
superpotential, K\"ahler potential and gauge kinetic functions \cite{inst1,inst2,inst3,inst4,inst5,Billo:2007sw,pablo}, with
possible applications to neutrino masses,  the $\mu$-term in MSSM, generation of perturbatively forbidden Yukawa
couplings in GUT's, and supersymmetry breaking. These instantonic
effects admit a nice geometrical description in
terms of Euclidean branes wrapping internal cycles. Therefore, the
very existence of rigid cycles, that can be wrapped
by the branes, has been argued of being of fundamental importance for the appearance
non-perturbative couplings in the superpotential,
since for these configurations the number of fermionic zero-modes can be appropriately reduced.
Along these lines, models with bulk branes intersecting instantons
wrapping rigid cycles \cite{rigidintersecting1} and
with D-branes wrapping partially rigid cycles
\cite{rigidintersecting2} have been constructed, while effects of
multi-instanton corrections have been analysed in  \cite{jacek}.

The goal of the present paper is twofold. First of all, we review
the construction of  $\mathbb {Z}_2 \times \mathbb {Z}_2$
orientifolds with fractional intersecting D-branes wrapping rigid
cycles first analysed in \cite{cristina,bcms}. We generalise the
construction in \cite{cristina}, and in particular we discuss in more detail the
emergence of new twisted tadpoles for RR four-forms. These tadpoles
are induced by the presence of the background magnetic field, and
can be explicitly computed from the one-loop vacuum amplitudes.
Alternatively, their emergence can also be derived from a study of
the low-energy Wess-Zumino couplings to twisted fields in the
presence of magnetic fields or from the cancelation of irreducible
non-Abelian anomalies. Secondly, we present consistent (global)
vacuum configurations with fractional D-branes wrapping rigid cycles, and we
study the role played by E1-branes wrapping the same cycles.
Instantonic effects can generate (model-dependent) superpotential
mass terms and/or linear terms for charged open-string fields, with
the possibility of breaking supersymmetry if one assumes that
closed-string moduli have been stabilised (as usual, supersymmetry
is restored if the closed string moduli are not stabilised and their
dynamics is taken into account). Along the way, we present a
concrete brane construction with ${\cal N}=1$ supersymmetry and
conformal invariance in the gauge sector at one-loop. One might even speculate,
along the lines of \cite{ls}, that the symmetry might still persist non-perturbatively.
Conformal
invariance is then broken by the instantonic effects at a
hierarchically small scale, thus providing\footnote{Our example is
however supersymmetric, whereas the proposal \cite{fv} was to
replace supersymmetry as a solution to the hierarchy problem by
conformal symmetry broken at low (TeV) scale. It is probably
possible to provide non-supersymmetric examples similar to the
present paper, possibly by analyzing the class of models in
\cite{aadds}.} a framework close to that of the Frampton-Vafa proposal for
the hierarchy problem \cite{fv}.

The paper is organised as follows. In Section 2, we review the
construction of the $\mathbb{Z}_2\times \mathbb{Z}_2$ orientifold
with discrete torsion and magnetised D-branes \cite{cristina,bcms},
generalising them along the lines of \cite{SSmagnetic}. In
particular, we pay special attention to new twisted RR tadpoles
induced by the presence of magnetic fields. We also give a detailed
analysis of the low-energy effective action and of the gauge anomaly
polynomial. In Section 3, we give an account of the charged zero
modes for the E1 instantons wrapping rigid cycles, and of their
non-perturbative contributions to the low-energy effective action.
Three globally consistent models are then explicitly constructed in
Section 4. The first model comprises only magnetised fractional D9
branes. The corresponding gauge theory is conformally invariant at
one-loop. However, instantons generate mass terms which break the
conformal symmetry at a hierarchically small scale.   In the second model, presented in Section 4.2, half of
the branes have been moved to bulk. In this class of models,
instantons wrapping rigid cycles generate mass terms mixing
open-string fields charged with respect to Abelian gauge fields on
the fractional branes together with open-string fields charged with
respect to Abelian gauge fields living on the bulk branes. In
Section 4.3 we build an orientifold model with gauge group
containing a $U(2)^4$ subgroup from magnetised branes. By the
introduction of suitable Wilson lines we prevent intersections of
the remaining unmagnetised branes with the instantonic Euclidean branes, and
therefore a linear term in the superpotential is generated for
charged fields in the antisymmetric representation of $U(2)_i$. These
non-perturbative contributions destabilise the vacuum and, assuming
the closed-string moduli are fixed to extrema of a flux-induced
potential, can potentially break supersymmetry. In all cases above,
we cross-check the instanton generated superpotential by the
gauge-invariance in the effective field theory. Section 5 contains our conclusions together with some speculations about the
fate of the non-supersymmetric model presented in \cite{aadds}.
Finally, Appendix A defines the characters used in this $\mathbb{Z}_2 \times \mathbb{Z}_2$ orbifold, while
Appendices B and C contain details on partition functions and spectra of magnetised D-branes and Euclidean-brane instanton.

\section{$\mathbb{Z}_2\times \mathbb{Z}_2$ orientifolds and discrete torsion}

We review here the construction of $\mathbb{Z}_2\times \mathbb{Z}_2$ orientifolds with discrete torsion
and magnetised D9-branes.  For simplicity, we consider a factorisable internal space
$T^6 = T^2_1 \times T^2_2 \times T^2_3$, and we denote by $o$, $g$, $f$ and $h$ the
four elements of $\mathbb{Z}_2\times \mathbb{Z}_2$, $o$ being the identity, acting
on the three internal two-tori as
\begin{equation}
g:(+,-,-) \ , \qquad f:(-,+,-) \ , \qquad h:(-,-,+) \ .
\end{equation}
A sign $\pm$ in the $i$-th position means that the two coordinates $(x^i,y^i)$ of the $T^2_i$ are mapped into
$\pm  (x^i,y^i)$ under the orbifold action.

As anticipated in the introduction, the one-loop partition function contains disconnected modular orbits involving twisted amplitudes, and therefore one has the freedom to introduce discrete torsion \cite{discretetorsion}. In this simple case, this freedom corresponds to a sign choice $\epsilon=\pm 1$ in front of the independent orbit. Clearly, the two choices correspond to different projections at the level of the spectrum, and indeed the Hodge numbers of the smooth Calabi-Yau manifolds associated to this $T^6/\mathbb{Z}_2 \times \mathbb{Z}_2$ orbifold are reversed.

In the case of the type IIB superstring, the one-loop torus amplitude is
\begin{equation}
\begin{split}
{\cal T}=&\frac{1}{4}\int_{\cal F}\frac{d^2\tau}{\tau_2^3}\Bigg\lbrace|T_{oo}|^2\varLambda_1\varLambda_2\varLambda_3
+|T_{og}|^2\varLambda_1\left|\frac{4\eta^2}{\theta_2^2}\right|^2
+|T_{of}|^2\varLambda_2\left|\frac{4\eta^2}{\theta_2^2}\right|^2
+|T_{oh}|^2\varLambda_3\left|\frac{4\eta^2}{\theta_2^2}\right|^2\\
&+|T_{go}|^2\varLambda_1\left|\frac{4\eta^2}{\theta_4^2}\right|^2
+|T_{gg}|^2\varLambda_1\left|\frac{4\eta^2}{\theta_3^2}\right|^2
+|T_{fo}|^2\varLambda_2\left|\frac{4\eta^2}{\theta_4^2}\right|^2
+|T_{ff}|^2\varLambda_2\left|\frac{4\eta^2}{\theta_3^2}\right|^2\\
&+|T_{ho}|^2\varLambda_3\left|\frac{4\eta^2}{\theta_4^2}\right|^2
+|T_{hh}|^2\varLambda_3\left|\frac{4\eta^2}{\theta_3^2}\right|^2\\
&+\epsilon\left(|T_{gh}|^2+|T_{gf}|^2+|T_{fg}|^2+|T_{fh}|^2
+|T_{hg}|^2+|T_{hf}|^2\right)\left|\frac{8\eta^3}{\theta_2\theta_3\theta_4}\right|^2\Bigg\rbrace
\frac{1}{|\eta|^2} \ ,
\end{split}\label{torus}
\end{equation}
where $\varLambda_k$ are lattice sums for the three internal
tori\footnote{For our notations and conventions, see \cite{emilian,carlo}.}. The
torus amplitude is expressed in terms of the $16$ quantities $T_{kl}$ given explicitly in Appendix A.
The sign $\epsilon=\mp 1$ multiplying the last line in the partition function, identifies the models with and without
discrete torsion. These have Hodge numbers $(h_{11}\, ,\, h_{21})=(3,51)$ and $(h_{11}\,,\,h_{21}) = (51,3)$, respectively, and are related by mirror symmetry.  On an orbifold, the Hodge homology is partially inherited from the covering space, $T^6$ in such case,
and  partially originates from the twisted sector. As we shall see, untwisted and twisted homology are intrinsically different and in fact induce completely different properties on D-branes.

Let us focus on the three cycles for this $T^6/\mathbb{Z}_2 \times \mathbb{Z}_2$ orbifold. Clearly, the number of inherited 3-cycles is $h_{21}^{\rm untw} = 8$, since these are the only 3-cycles of the covering torus that survive the orbifold projection, and are insensitive to discrete torsion. On the contrary, twisted three-cycles only exist in the case discrete torsion is turned on, and more importantly their topology is $S^2 \times S^1$, where $S^2$ corresponds to the two-cycle of the blown-up singularity and $S^1$ is a one-cycle of the fixed torus. Now, a 3-cycle wrapping bulk and twisted cycles, referred to in the literature as rigid cycle, is localised at orbifold fixed points and cannot be deformed. As a result, if D-branes are wrapped along these rigid cycles the translational modes in the compact directions should be absent, and therefore no adjoint scalars (and extra fermions) appear in the spectrum\footnote{For a more detailed description of the homology of the $T^6/\mathbb{Z}_2 \times \mathbb{Z}_2$ orbifold and the role of rigid cycles in orientifold constructions see \cite{bcms}.}. That's why in the following we shall confine our attention to the $\mathbb{Z}_2 \times \mathbb{Z}_2$ orbifold with discrete torsion. In the literature these are also known as $T^6/\mathbb{Z}_2 \times \mathbb{Z}_2'$ orbifolds, to distinguish them from the $T^6/\mathbb{Z}_2 \times \mathbb{Z}_2$ orbifolds without discrete torsion.

We can now mod-out the IIB superstring by the world-sheet parity $\varOmega$ that, as usual, is implemented by the
Klein-bottle amplitude
\begin{equation}
\begin{split}
{\cal K}=&\frac{1}{4}\int_0^{\infty}\frac{d\tau_2}{\tau_2^3} \frac{1}{\eta^2} \,\Bigg\lbrace
\left(P_1P_2P_3+P_1W_2W_3+W_1P_2W_3+W_1W_2P_3\right)T_{oo}
\\
&+2\times16\, \left[\epsilon_1(P_1+\epsilon W_1)T_{go}+\epsilon_2(P_2+\epsilon W_2)T_{fo}+
\epsilon_3(P_3+\epsilon
W_3)T_{ho}\right]\left(\frac{\eta}{\theta_4}\right)^2\Bigg\rbrace
\ , \\
\end{split}\label{klein}
\end{equation}
so that the spectrum of the unoriented closed strings on the $T^6 /\mathbb{Z}_2 \times \mathbb{Z}_2$ orbifold is encoded in
\begin{equation}
\frac{1}{2} \left( {\cal T} + {\cal K} \right) \,.
\end{equation}
In eq.~(\ref{klein}), $P_k$ and $W_k$ denote the restriction of the
lattice sums $\varLambda_k$ to their momentum and winding
sublattices. The option of turning on discrete torsion in
eq.~(\ref{torus}) affects the Klein-bottle amplitude by  the
presence of the three signs $\epsilon_k =\pm$, that must obey
\begin{equation}\label{epsilon}
\epsilon_1 \ \epsilon_2 \ \epsilon_3 \ = \ \epsilon \, .
\end{equation}
Notice, in particular, the effect of discrete torsion on
$P_k+\epsilon W_k$. If $\epsilon=1$, the type IIB torus amplitude
eq.~(\ref{torus}) corresponds to the so-called diagonal modular
invariant partition function and, as a results, all states,
including the twisted ones, must be properly projected by the
Klein-bottle amplitude, as is indeed the case. On the contrary, if
discrete torsion is turned on and  $\epsilon=-1$, the type IIB torus
amplitude eq.~(\ref{torus}) corresponds to the so-called
charge-conjugation modular invariant partition function. Since the
twisted characters are all complex, the twisted
amplitudes involve only off-diagonal combinations of holomorphic and
anti-holomorphic characters. As a result, $\varOmega$ has a trivial
action on them, and they cannot contribute to ${\cal K}$. This is
clearly spelled in the combinations $P_k+\epsilon W_k = P_k - W_k$.

An $S$ modular transformation brings ${\cal K}$ into the corresponding tree-level (transverse) channel:
\begin{equation}
\begin{split}
\tilde{\cal K}=&
\frac{2^5}{4}\Bigg\lbrace\left(v_1v_2v_3W_1^eW_2^eW_3^e+\frac{v_1}{v_2v_3}W_1^eP_2^eP_3^e
+\frac{v_2}{v_1v_3}P_1^eW_2^eP_3^e+\frac{v_3}{v_1v_2}P_1^eP_2^eW_3^e\right)T_{oo}
\\
& +2\left[\epsilon_1\left(v_1W_1^e+\epsilon\frac{P_1^e}{v_1}\right)T_{og}
+\epsilon_2\left(v_2W_2^e+\epsilon\frac{P_2^e}{v_2}\right)T_{of} \right.
\\
&\left.+\epsilon_3\left(v_3W_3^e+\epsilon\frac{P_3^e}{v_3}\right)T_{oh}\right]\left(\frac{2\eta}{\theta_2}\right)^2\Bigg\rbrace \,,
\\
\end{split}
\end{equation}
where the superscript $e$ stands for the usual restriction of the lattice terms to their even subsets.
As expected, the massless states group into perfect squares
\begin{equation}
\begin{split}
\tilde{\cal K}_0 =&
\frac{2^5}{4}\Bigg\lbrace\left(\sqrt{v_1v_2v_3}+\epsilon_1\sqrt{\frac{v_1}{v_2v_3}}
+\epsilon_2\sqrt{\frac{v_2}{v_1v_3}}+\epsilon_3\sqrt{\frac{v_3}{v_1v_2}}\right)^2\tau_{oo}
\\
&+\left(\sqrt{v_1v_2v_3}+\epsilon_1\sqrt{\frac{v_1}{v_2v_3}}
-\epsilon_2\sqrt{\frac{v_2}{v_1v_3}}-\epsilon_3\sqrt{\frac{v_3}{v_1v_2}}\right)^2\tau_{og}
\\
&+\left(\sqrt{v_1v_2v_3}-\epsilon_1\sqrt{\frac{v_1}{v_2v_3}}
+\epsilon_2\sqrt{\frac{v_2}{v_1v_3}}-\epsilon_3\sqrt{\frac{v_3}{v_1v_2}}\right)^2\tau_{of}
\\
&+\left(\sqrt{v_1v_2v_3}-\epsilon_1\sqrt{\frac{v_1}{v_2v_3}}
-\epsilon_2\sqrt{\frac{v_2}{v_1v_3}}+\epsilon_3\sqrt{\frac{v_3}{v_1v_2}}\right)^2\tau_{oh}\Bigg\rbrace \,.
\\
\end{split}
\label{kleinzero}
\end{equation}
The coefficients of the untwisted $\tau_{ok}$ characters\footnote{For the definition of the $\tau_{kl}$ characters see \cite{massimo} and \cite{carlo}.} clearly spell-out the geometry of the orientifold planes involved by this $T^6 /\mathbb{Z}_2 \times \mathbb{Z}_2$ orientifold, that crucially depends on the choice of the sign $\epsilon$. In all cases, the ubiquitous O9 planes are accompanied by three different families of O5 planes, each wrapping one of the internal $T^2$'s. However, if on the one hand models {\it without} discrete torsion ($\epsilon =+1$) can consistently accommodate ``conventional'' ${\rm O}_-$ planes with negative tension and charge, on the other hand models {\it with} discrete torsion ($\epsilon =-1$) must involve at least one ``exotic'' ${\rm O}5_+$ plane with positive tension and charge. In the following, we shall confine our attention to the class of models with discrete torsion corresponding to the choice $(\epsilon_1,\epsilon_2,\epsilon_3)=(+,+,-)$.

We can now turn to the open-string sector. The simplest way to
cancel the RR tadpoles from eq.~(\ref{kleinzero}) is to introduce
suitable numbers of D9 branes, of ${\rm D}5_1$ branes wrapping the
first $T^2$, of ${\rm D}5_2$ branes wrapping the second $T^2$ and,
finally, of $\overline{{\rm D}5}_3$ branes wrapping the third $T^2$
\cite{aadds}. The anti-branes are needed to cancel the positive
charge of the corresponding ${\rm O}5_+$ plane. With this
configuration of O-planes and D-branes, NS-NS tadpoles cannot be
cancelled and supersymmetry is explicitly broken as in \cite{BSB}, resulting in nonsupersymmetric but tachyon-free models.
However, it has been shown in \cite{ms,cristina,bcms} that
supersymmetric vacua can be actually built if suitable background
magnetic fields are turned on.

A proper implementation of the orbifold projection in the
open-string sector, suggests that we introduce two ``families'' of
D9-branes, labelled by the complex charges $p_a$, $\bar p_a$,
$q_\alpha$ and $\bar q_\alpha$, so that the action of $\mathbb{Z}_2
\times \mathbb{Z}_2$ on the Chan-Paton labels is
\begin{equation}
\begin{split}
&N_{a,o}=p_a+\bar p_a \,,
\\
&N_{a,g}=i(p_a-\bar p_a) \,,
\\
&N_{a,f}=i(p_a-\bar p_a)\,,
\\
&N_{a,h}=p_a+\bar p_a\,,
\\
\end{split}\qquad
\begin{split}
&N_{\alpha,o}=q_{\alpha}+\bar q_{\alpha} \,,
\\
&N_{\alpha,g}=i(q_{\alpha}-\bar q_{\alpha})\,,
\\
&N_{\alpha,f}=-i(q_{\alpha}-\bar q_{\alpha})\,,
\\
&N_{\alpha,h}=-q_{\alpha}-\bar q_{\alpha} \,,
\\
\end{split}
\label{cp}
\end{equation}
and the resulting gauge group is the product of unitary factors
\begin{equation}
G_{\rm CP} = \prod_aU(p_a)\times\prod_{\alpha} U(q_{\alpha}) \,.
\end{equation}
Since the parametrisation in eqs.~(\ref{cp}) determines the D-brane
contribution to the untwisted ($N_{a,o}$ and $N_{\alpha ,o}$) and
$k$-twisted ($N_{a,k}$ and $N_{\alpha ,k}$) tadpoles, it is clear
that, before magnetic backgrounds are turned on, D9 branes have
non-trivial couplings only with respect to forms from the
$h$-twisted sector. This is a consequence of the presence of the
${\rm O}5_+$ planes wrapping the third torus fixed under the $h$
generator.

Turning on magnetic backgrounds for the open-string gauge fields
affects as usual the masses of the open-string modes and the
multiplicities of the various representations. Details on the
structure of annulus and M\"obius-strip partition functions for the
case of constant magnetic fields $H^{(A)}_i$ on the $A$-th stack of
D9 branes where $A = (a,\alpha)$ along the $T^2_i$ torus are given
in Appendix B. Dirac quantisation condition
\begin{equation}
H_i^{(A)} \  = \ \frac{m_i^{(A)}}{n_i^{(A)} v_i} \ ,
\end{equation}
relates the volume $v_i$ of the torus to the strength of the magnetic
deformation, while demanding that $\mathcal{N}=1$ supersymmetry be
preserved in four dimensions constrains the choices of the magnetic
fields along the $T^2$'s
\begin{equation}
H_1^{(A)} + H_2^{(A)} + H_3^{(A)} \  = \  H_1^{(A)} H_2^{(A)}
H_3^{(A)} \, .
\label{SUSY}
\end{equation}
With a T-duality along the horizontal directions of the three two-torii, D9-branes are transformed into D6 branes, and the magnetic fields $H^{(A)}_i$ are related to the angles $\theta_i^{(A)}$ the rotated D6-branes make with the horizontal axis of the $i$-th torus, $H^{(A)}_i = \tan\, \theta_i^{(A)}$.  The integers $(m_i^{(A)},n_i^{(A)})$, that in the magnetised picture denote the familiar degeneracies of the Landau levels from quantum mechanics, now count the number of times D6 branes wrap the fundamental cycles of the $T^2_i$, while the supersymmetry condition eq. (\ref{SUSY}) now becomes
\begin{equation}
\theta_1^{(A)} + \theta_2^{(A)} + \theta_3^{(A)} = 0\,.
\end{equation}
Although the condition in eq. (\ref{SUSY}) guarantees that the $A$-th stack of magnetised branes preserves $\mathcal{N}=1$ supersymmetry, demanding that all stacks preserve the same supersymmetry charges further constrains the magnetic fields to satisfy the inequality
\begin{equation}
H_1^{(A)} H_2^{(A)} + H_1^{(A)} H_3^{(A)} + H_2^{(A)} H_3^{(A)} \ \leq
1 \,.
\label{Tadpole_SUSY}
\end{equation}
Therefore, if and only if eqs.~(\ref{SUSY}) and (\ref{Tadpole_SUSY}) hold for every stack $A$, the resulting vacuum configuration is supersymmetric.

In addition to the supersymmetry conditions, consistent models must also satisfy tadpole conditions, for both RR and NS-NS massless states. These can be divided into two classes, untwisted and twisted, depending on the origin of the corresponding closed-string fields.  From eq. (\ref{kleinzero}) and from the annulus and M\"obius strip amplitudes in the Appendix, one can derive the conditions
\begin{equation}
\begin{split}
\sum_a p_a \, n_1^{(a)}n_2^{(a)}n_3^{(a)}+\sum_{\alpha} q_{\alpha} \, n_1^{(\alpha)}n_2^{(\alpha)}n_3^{(\alpha)} &=16 \,,
\\
\sum_a p_a \, n_1^{(a)}m_2^{(a)}m_3^{(a)}+\sum_{\alpha} q_{\alpha} \, n_1^{(\alpha)}m_2^{(\alpha)}m_3^{(\alpha)} &=-16\, \epsilon_1\,,
\\
\sum_a p_a \, m_1^{(a)}n_2^{(a)}m_3^{(a)}+\sum_{\alpha} q_{\alpha} \, m_1^{(\alpha)}n_2^{(\alpha)}m_3^{(\alpha)} &=-16 \, \epsilon_2\,,
\\
\sum_a p_a \, m_1^{(a)}m_2^{(a)}n_3^{(a)}+\sum_{\alpha} q_{\alpha} \, m_1^{(\alpha)}m_2^{(\alpha)}n_3^{(\alpha)} &=-16 \, \epsilon_3\,,
\\
\end{split} \label{untwisted}
\end{equation}
from the untwisted sector, and
\begin{equation}
\begin{split}
\sum_a p_a \, m_1^{(a)}\epsilon_l^{(a),g}+\sum_{\alpha} q_{\alpha}\, m_1^{(\alpha)}\epsilon_l^{(\alpha),g} &= \ 0 \,,
\\
\sum_a p_a \, m_2^{(a)}\epsilon_l^{(a),f}-\sum_{\alpha} q_{\alpha} \, m_2^{(\alpha)}\epsilon_l^{(\alpha),f} & = \ 0 \,,
\\
\sum_a p_a \, n_3^{(a)}\epsilon_l^{(a),h}-\sum_{\alpha} q_{\alpha} \, n_3^{(\alpha)}\epsilon_l^{(\alpha),h} & = \ 0 \,,
\\
\end{split} \label{twisted}
\end{equation}
for each twisted sector $g$, $f$ and $h$. Clearly, twisted tadpoles must be satisfied locally at each fixed point $l=1,\ldots , 16$. Twisted tadpole conditions have a neat geometrical interpretation in the T-dual picture in terms of branes at angles \cite{SSmagnetic}.  Since the twisted closed-string fields are localised at the fixed point, only branes passing through the $l$-th fixed point can in principle couple to them and contribute to their tadpole. In eq. (\ref{twisted}), this is neatly spelled by the  index
$\epsilon_l^{(A),g}$ that is equal to $1$ if brane $A$ passes through the $l$-th point fixed under the action of the $g$ generator, and equals zero otherwise.

Notice the deep difference in the brane contribution to the twisted tadpoles in eq.~(\ref{twisted}), as a result of
the type of O-planes involved and the consequent action of the $\mathbb{Z}_2 \times \mathbb{Z}_2$ orbifold on the Chan-Paton labels, eq.~(\ref{cp}). In the $g$ and $f$ twisted sectors, branes couple to the corresponding RR forms only in the presence of a non-trivial magnetic field
($m^{(A)}_i \not = 0$). For instance, if one considers the coupling to the RR field $S_2C_2O_2O_2$ from the $g$-twisted sector, one finds
\begin{equation}
\begin{split}
&\frac{n_1^{(a)}}{2}\, \epsilon_l^{(a),g}\, \left[p_a\int_{\mathcal{M}_6}C^{(l)}\wedge
e^{s_1H_1^{(a)}}-\bar{p}_a\int_{\mathcal{M}_6}C^{(l)}\wedge
e^{-s_1H_1^{(a)}}\right]
\\
& +\frac{n_1^{(\alpha)}}{2}\, \epsilon_l^{(\alpha),g}\, \left[q_{\alpha}\int_{\mathcal{M}_6}C^{(l)}\wedge
e^{s_1H_1^{(\alpha)}}-\bar{q}_{\alpha}\int_{\mathcal{M}_6}C^{(l)}\wedge
e^{-s_1H_1^{(\alpha)}}\right]
\\
&=\frac{s_1}{2}\left[(p_a+\bar p_a)m_1^{(a)}\epsilon_l^{(a),g}
+(q_{\alpha}+\bar
q_{\alpha})m_1^{(\alpha)}\epsilon_l^{(\alpha),g}\right]\int_{\mathcal{M}_4}C^{(l)}_4 \,,
\\
\end{split}
\label{twistedcoupling1}
\end{equation}
where $s_i = \pm 1$ represent the internal helicities of the fermions and $C_j^{(l)}$ are $j$-form RR fields localised at the $l$-th fixed point.

On the contrary, D9 branes couple to the $h$-twisted RR fields also if the magnetic field is turned off ({\it i.e.} if $m_3 = 0$ and $n_3 =1$). In this case, in the low-energy effective action one finds couplings of the form
\begin{equation}
\begin{split}
&\frac{n_3^{(a)}}{2}\, \epsilon_l^{(a),h}\, \left[p_a\int_{\mathcal{M}_6}C^{(l)}\wedge
e^{s_3H_3^{(a)}}+\bar{p}_a\int_{\mathcal{M}_6}C^{(l)}\wedge
e^{-s_3H_3^{(a)}}\right]
\\
&-\frac{n_3^{(\alpha)}}{2}\, \epsilon_l^{(\alpha),h}\, \left[q_{\alpha}\int_{\mathcal{M}_6}C^{(l)}\wedge
e^{s_3 H_3^{(\alpha)}}+\bar{q}_{\alpha}\int_{\mathcal{M}_6}C^{(l)}\wedge
e^{-s_3 H_3^{(\alpha)}}\right]
\\
&=\frac{1}{2}\left[\epsilon_l^{(a),h} n_3^{(a)} \left(p_a + \bar{p}_a\right) -\epsilon_l^{(\alpha),h} n_3^{(\alpha)} \left(q_{\alpha}+\bar{q}_{\alpha}\right)\right]\int_{\mathcal{M}_6}C_6^{(l)} \,.
\\
\end{split}
\label{twistedcoupling2}
\end{equation}

In addition to the cancellation of homological tadpoles, that have a clear low-engery descriptions in terms of the consistency of Bianchi identities and equations of motion, one has to impose further K-theory constraints, associated to some torsion charges which are invisible to homology \cite{Ktheory}. For such $T^6 / \mathbb{Z}_2 \times \mathbb{Z}_2$ orientifold, K-theory conditions were derived in \cite{bcms}, and we have explicitly checked that, in all the examples we shall discuss in this paper, they are satisfied.

The magnetic field configuration and/or the geometry of the D-branes at angles also affects the massless excitations in the open-string sector. If the supersymmetry conditions eqs.~(\ref{SUSY}) and (\ref{Tadpole_SUSY}) are satisfied, the light fields comprise chiral superfields charged with respect to the gauge group $G_{\rm CP}$ as listed in table \ref{table1}.

\begin{table}[htdp]
\begin{center}
\begin{tabular}{|c|c|c|}
\hline
Multiplicity & Representation & Relevant Indices \\
\hline
\hline
$\frac{1}{8}(I^{aa'}+I^{aO}-4I_1^{aa'}-4I_2^{aa'}+4I_3^{aa'})$ & $\left(\frac{p_a(p_a-1)}{2},1\right)$ & $\forall a$
\\[1ex]
$\frac{1}{8}(I^{\alpha\alpha'}+I^{\alpha O}-4I_1^{\alpha\alpha'}-4I_2^{\alpha\alpha'}+4I_3^{\alpha\alpha'})$ &
$\left(1,\frac{q_{\alpha}(q_{\alpha}-1)}{2}\right) $&$\forall \alpha$
\\[1ex]
$\frac{1}{8}(I^{aa'}-I^{aO}-4I_1^{aa'}-4I_2^{aa'}+4I_3^{aa'})$ & $\left(\frac{p_a(p_a+1)}{2},1\right)$ & $\forall a$
\\[1ex]
$\frac{1}{8}(I^{\alpha\alpha'}-I^{\alpha O}-4I_1^{\alpha\alpha'}-4I_2^{\alpha\alpha'}+4I_3^{\alpha\alpha'})$ &
$\left(1,\frac{q_{\alpha}(q_{\alpha}+1)}{2}\right) $&$\forall \alpha$
\\[1ex]
\hline
$\frac{1}{4}(I^{a\alpha'}-S_g^{a\alpha}I_1^{a\alpha'}+S_f^{a\alpha}I_2^{a\alpha'}-S_h^{a\alpha}I_3^{a\alpha'})$& $(p_a,q_{\alpha})$ & $\forall a,\forall\alpha$
\\[1ex]
$\frac{1}{4}(I^{a\alpha}+S_g^{a\alpha}I_1^{a\alpha}-S_f^{a\alpha}I_2^{a\alpha}-
S_h^{a\alpha}I_3^{a\alpha})$& $(p_a,\bar{q}_{\alpha})$ &$\forall a,\forall \alpha$
\\[1ex]
\hline
$\frac{1}{4}(I^{ab'}-S_g^{ab}I_1^{ab'}-S_f^{ab}I_2^{ab'}+S_h^{ab}I_3^{ab'})$ & $(p_a,p_b)$ & $a < b$
\\[1ex]
$\frac{1}{4}(I^{ab}+S_g^{ab}I_1^{ab}+S_f^{ab}I_2^{ab}+S_h^{ab}I_3^{ab})$& $(p_a,\bar{p}_b)$& $a < b$
\\[1ex]
$\frac{1}{4}(I^{\alpha\beta'}-S_g^{\alpha\beta}I_1^{\alpha\beta'}-S_f^{\alpha\beta}I_2^{\alpha\beta'}
+S_h^{\alpha\beta}I_3^{\alpha\beta'})$ & $(q_{\alpha},q_{\beta})$& $\alpha < \beta$
\\[1ex]
$\frac{1}{4}(I^{\alpha\beta}+S_g^{\alpha\beta}I_1^{\alpha\beta}+S_f^{\alpha\beta}I_2^{\alpha\beta}
+S_h^{\alpha\beta}I_3^{\alpha\beta})$&  $(q_{\alpha},\bar{q}_{\beta})$ & $\alpha < \beta $
\\[1ex]
\hline
1 & $(p_a,\bar{q}_\alpha) + (\bar{p}_a, q_\alpha)$ & if $H_i^a = H_i^\alpha~\forall i$
\\[1ex]
\hline
1 & $(p_a,q_\alpha) + (\bar{p}_a, \bar{q}_\alpha)$ & if $H_i^a = - H_i^\alpha~\forall i$
\\[1ex]
\hline
\end{tabular}
\end{center}
\caption{Representations and multiplicities of charged chiral
superfields on a $T^6 /\mathbb{Z}_2 \times \mathbb{Z}_2$ orbifold
with discrete torsion, in the presence of magnetic backgrounds.}
\label{table1}
\end{table}%

The effective multiplicities of representations for this
$T^6/\mathbb{Z}_2 \times \mathbb{Z}_2$ orbifold, depend not only on
the topological intersection numbers between branes of different
types
\begin{equation}
\begin{split}
I^{AB} &=\prod_{i=1}^3I_i^{AB} \,, \qquad I_i^{AB} = m_i^{(A)}n_i^{(B)}-n_i^{(A)}m_i^{(B)} \,,
\\
I^{AB'} &=\prod_{i=1}^3I_i^{AB'} \,, \qquad I_i^{AB'} = m_i^{(A)}n_i^{(B)}+n_i^{(A)}m_i^{(B)} \,,
\\
\end{split}
\label{intersections}
\end{equation}
and branes and O-planes
\begin{equation}
I^{AO}=8\left( m_1^{(A)}m_2^{(A)}m_3^{(A)}-\epsilon_1\, m_1^{(A)}n_2^{(A)}n_3^{(A)}-\epsilon_2\, n_1^{(A)}m_2^{(A)}n_3^{(A)}
-\epsilon_3\, n_1^{(A)}n_2^{(A)}m_3^{(A)} \right),
\label{ointersections}
\end{equation}
but they also depend on the action of the orbifold group on the
open-string states and the Chan-Paton factors. Since the orbifold
projection acts through the fixed points, it is clear that states at
brane intersections feel the orbifold projection if and only if the
intersection point coincides with one of the orbifold fixed points
\cite{SSmagnetic}. This is encoded in the quantity
\begin{equation}
\begin{split}
S_{i=g,f,h}^{AB}= &\  {\rm number\ of\ fixed\ points\ of\ the\
generator\ } i=g,f,h \
\\
&\  {\rm that\ both\ branes\ } A \ {\rm and}\ B \ {\rm intersect}\,,
\\
\end{split}
\label{fixedintersections}
\end{equation}
with $A=(a,\alpha)$, $B=(b,\beta)$. Notice that, under the working
assumption that all branes pass through the origin of the three
two-tori, $S_{i=g,f,h}^{aa}=S_{i=g,f,h}^{\alpha\alpha}=4$,  as is
explicitly written in table \ref{table1}. $S_{i}^{AB}$ depends on
the wrapping numbers of the two stacks $A$ and $B$. More
specifically, for $\mathbb{Z}_2$ fixed points they only depend on
the equivalence class branes $A$ and branes $B$ belong to. In
particular, if they belong to the same equivalence class ({\it i.e.}
the corresponding wrapping numbers have the same parity) then
they intersect once or twice at the fixed points in a given torus.
In our case, since the
$T^6/\mathbb{Z}_2 \times \mathbb{Z}_2$ orbifold, strictly speaking,
does not admit fixed points but only fixed two-tori, $S^{AB}_i$ can
only take the values 1, 2 or 4. This is crucial in order to give a
consistent spectrum with integer multiplicities, as can be deduced
from table \ref{table1}.

As usual, tadpole conditions take care of irreducible anomalies,
whenever present. In fact, in the case at hand, the chiral spectrum
in table \ref{table1} is potentially anomalous, with a coefficient
of the $SU(p_a)^3$ irreducible anomaly proportional to
\begin{equation}
\begin{split}
SU(p_a)^3 \sim &\  m_1^{(a)}m_2^{(a)}m_3^{(a)}\left[\sum_b
p_b \, n_1^{(b)}n_2^{(b)}n_3^{(b)}+\sum_{\alpha} q_{\alpha} \, n_1^{(\alpha)}n_2^{(\alpha)}n_3^{(\alpha)}-16\right]
\\
&+ m_1^{(a)}n_2^{(a)}n_3^{(a)}\left[\sum_b
p_b \, n_1^{(b)}m_2^{(b)}m_3^{(b)}+\sum_{\alpha} q_{\alpha}\, n_1^{(\alpha)}m_2^{(\alpha)}m_3^{(\alpha)}+16\, \epsilon_1\right]
\\
&+ n_1^{(a)}m_2^{(a)}n_3^{(a)}\left[\sum_b
p_b\, m_1^{(b)}n_2^{(b)}m_3^{(b)}+\sum_{\alpha} q_{\alpha}\, m_1^{(\alpha)}n_2^{(\alpha)}m_3^{(\alpha)}+16\,\epsilon_2\right]
\\
&+ n_1^{(a)}n_2^{(a)}m_3^{(a)}\left[\sum_b
p_b\, m_1^{(b)}m_2^{(b)}n_3^{(b)}+\sum_{\alpha} q_{\alpha}\, m_1^{(\alpha)}m_2^{(\alpha)}n_3^{(\alpha)}+16\,\epsilon_3\right]
\\
&- n_1^{(a)}\left[\sum_b
S_g^{ab}\, p_b \, m_1^{(b)}+\sum_{\alpha} S_g^{a\alpha}\, q_{\alpha}\, m_1^{(\alpha)}\right]
\\
&- n_2^{(a)}\left[\sum_b
S_f^{ab}\, p_b \, m_2^{(b)}+\sum_{\alpha} S_f^{a\alpha}\, q_{\alpha}\, m_2^{(\alpha)}\right]
\\
&+m_3^{(a)}\left[\sum_b S_h^{ab}\, p_b \, n_3^{(b)}+\sum_{\alpha} S_h^{a\alpha}\, q_{\alpha}\, n_3^{(\alpha)}\right] \,,
\\
\end{split}
\label{anomaly}
\end{equation}
where the sums extend over all stacks of branes. Clearly, each line
is proportional to the tadpoles in eqs. (\ref{untwisted}) and
(\ref{twisted}) and, therefore, the anomaly is cancelled in
consistent vacuum configurations, where tadpoles are absent. Notice
that, in order to recover the twisted tadpoles in eq.
(\ref{anomaly}), we have explicitly made use of the relation
\begin{equation}
S_k^{ab}=\sum_{l\in F_k}\epsilon_l^{(a),k}\, \epsilon_l^{(b),k}\,,
\qquad {\rm for}\quad k=g,h,f \,.
\end{equation}
where in the sum above, $F_k$ denotes the set of fixed points of the
operation $k=g,h,f$. Similar expressions clearly hold for $SU(q_\alpha )^3$, and other combinations of unitary groups.

\subsection{The low-energy effective action}

From the tadpole conditions and from eqs. (\ref{twistedcoupling1})
and (\ref{twistedcoupling2}), one can easily deduce some of the
couplings in the low-energy effective action. In particular, if one
denotes by $S$ the axion-dilaton chiral superfield, by $T_i$ and
$U_i$, $i=1,2,3$, the K\"ahler and complex structure moduli associated
to the three two-tori, and by $M_{i}^{l}$, $l=1,\ldots ,16$, the
$48=3\times 16$ chiral multiplets from the twisted sectors, one
for each fixed point, the gauge kinetic functions for the magnetised
D9 branes read
\begin{equation}
\begin{split}
f_{D9}^{(A)}=
&n_1^{(A)}n_2^{(A)}n_3^{(A)}\, S - n_1^{(A)}m_2^{(A)}m_3^{(A)}\, T_1 - m_1^{(A)}n_2^{(A)}m_3^{(A)} \, T_2 -m_1^{(A)}m_2^{(A)}n_3^{(A)}\, T_3
\\
&+\alpha_1 \left( \sum_{l\in F_g} \epsilon_l^{(A),g}\,
X_1^{(A)}\, m_1^{(A)}\, M_1^l
+\sum_{l\in F_f} \epsilon_l^{(A),f}\, X_2^{(A)}\, m_2^{(A)}\, M_2^l \right)
\\
&+ \alpha_2\, \sum_{l\in F_h} \epsilon_l^{(A),h}\, X_3^{(A)}\, n_3^{(A)} \, M_3^h \,.
\end{split}
\end{equation}
Clearly, $A=(a,\alpha)$ counts the different type of magnetised
branes, $\alpha_1$ and $\alpha_2$ are constants,
and $M_i^l$ denotes the twisted fields
localised at the $l$-th fixed point of type $i$ where the ${\rm D}9^{(A)}$ brane
passes through. Finally, the $X_i^{(A)}$'s represent the ``charge''
of the twisted moduli with respect to the gauge groups $U(p_a)$
relative to $U(q_{\alpha})$. They can be extracted from eq.
(\ref{twisted}), and
\begin{equation}
\begin{split}
&X_1^{(a)}=+1 \, , \qquad X_1^{(\alpha)}=+1
\\
&X_2^{(a)}=+1 \, , \qquad X_2^{(\alpha)}=-1
\\
&X_3^{(a)}=+1 \, , \qquad X_3^{(\alpha)}=-1 \ ,
\\
\end{split}
\end{equation}
if the $A$ brane passes through the given fixed points, and vanish otherwise.

The closed-string fields also transform with respect to the
$U(1)^{(A)}$ gauge transformations $\varLambda^{(A)}$, as demanded
by the generalised Green-Schwarz mechanism. If the Abelian factor is
associated to a stack of $p_A= \{ p_a, q_\alpha\}$
D-branes, the non-linear transformations of the closed-string fields
read
\begin{equation}
\begin{split}
&\delta S=\lambda_S\sum_A p_A~ m_1^{(A)}m_2^{(A)} m_3^{(A)} \varLambda^{(A)} \,,
\\
&\delta T_1=\lambda\sum_A p_A~ m_1^{(A)}n_2^{(A)} n_3^{(A)} \varLambda^{(A)} \,,
\\
&\delta T_2=\lambda\sum_A p_A~ n_1^{(A)}m_2^{(A)} n_3^{(A)} \varLambda^{(A)} \,,
\\
&\delta T_3=\lambda\sum_A p_A~ n_1^{(A)}n_2^{(A)} m_3^{(A)} \varLambda^{(A)} \,,
\\
&\delta M_1=\lambda_1\, \alpha_1 \sum_A p_A ~\epsilon_l^{(A),g}~X_1^{(A)} n_1^{(A)} \varLambda^{(A)} \,,
\\
&\delta M_2=\lambda_2\, \alpha_1\sum_A p_A~ \epsilon_l^{(A),f}~X_2^{(A)} n_2^{(A)} \varLambda^{(A)} \,,
\\
&\delta M_3=\lambda_3\, \alpha_2\sum_A p_A~ \epsilon_l^{(A),h}~X_3^{(A)} m_3^{(A)} \varLambda^{(A)} \,,
\\
\end{split}
\label{variation4}
\end{equation}
where the $\lambda$'s are normalisation constants, that can be straightforwardly determined by comparing a generic $U(1)$ anomaly matrix
\begin{equation}
C_{AB}= \frac{1}{4\pi^2}{\rm Tr}\, (Q_A^2Q_B) \, , \qquad C_{AA}= \frac{1}{12 \pi^2}\, {\rm Tr}\, (Q_A^3) \,,
\label{mixedanom}
\end{equation}
with the variation of the gauge kinetic function
\begin{equation}
\delta f^A=\sum_B C_{AB}\, \varLambda^{(B)}
\end{equation}
under generic Abelian transformations $V^{(A)} \rightarrow V^{(A)} +
\varLambda^{(A)} + {\bar \varLambda}^{(A)} $. In our case, however,
one has to be careful since the generators of the $U(1)$'s are not
canonically normalised, so that an additional ${\rm Tr}\,  Q_A^2$
factor appears,
\begin{equation}
\delta {\cal L} =  \sum_{A,B} C_{AB} \, \varLambda^{(B)}\, {\rm Tr}
( Q_A^2)\, F_A \, \tilde{F}_A \,.
\end{equation}

The gauge variations eq.~(\ref{variation4}) lead to the appearance of FI terms
$\xi^{(A)}$ on the magnetised stacks. In the orbifold limit $M_i \ll
1$, they are given by
\begin{equation}
\begin{split}
\frac{1}{p_A} \xi^{(A)}  \ =& \ {\lambda_S \, m_1^{(A)} m_2^{(A)} m_3^{(A)}
  \over s} + \lambda \left( {m_1^{(A)} n_2^{(A)} n_3^{(A)} \over t_1} + {n_1^{(A)} m_2^{(A)}
  n_3^{(A)} \over t_2} + {n_1^{(A)} n_2^{(A)} m_3^{(A)} \over t_3}
\right)
\\
& - \alpha_1 \, \left( \lambda_1\, \sum_{l\in F_g} \epsilon_l^{(A),g}~
 {\cal M}_1^l \, X_1^{(A)} \, n_1^{(A)} + \lambda_2 \, \sum_{l\in F_f} \epsilon_l^{(A),f}~
{\cal M}_2^l \, X_2^{(A)} \, n_2^{(A)} \right)
\\
&- \lambda_3 \, \alpha_2 \,\sum_{l\in F_h} \epsilon_l^{(A),h}~ {\cal M}_3^l \, X_3^{(A)} \,
m_3^{(A)} \,,
\end{split}
\label{fi}
\end{equation}
where we defined the real parts  $s = {\rm Re} \, S$,
$t_i = {\rm Re} \, T_i$ and  ${\cal M}_i = {\rm Re} \, M_i$ of the closed-string moduli.

\section{Rigid instantons}

Instantonic corrections to the low-energy effective action can also
be computed in terms of Euclidean branes fully wrapping internal
cycles. We postpone until Appendix C details about the one-loop
amplitudes encoding the spectra of the E1 and E5 instantons on this
$\mathbb{Z}_2 \times \mathbb{Z}_2$ orientifold with discrete
torsion, and with $(\epsilon_1,\epsilon_2,\epsilon_3) = (1,1, -1)$.
There it is shown that only  ${\rm E}1_3$ instantons, wrapping the
same rigid cycle as the ${\rm O}5_+$ planes, have the required
minimal number (two) of fermionic zero modes to generate directly
non-perturbative contributions to the superpotential. ${\rm E}1_1$,
${\rm E}1_2$ and ${\rm E}5$ instantons have instead four neutral
zero-modes and, therefore, they might yield non-trivial
contributions to the superpotential only if they properly interact
with zero-modes on different branes \cite{petersson}. However, we
shall not discuss here this possibility, and consider only the
effect of a single E1$_3$ instanton, also known as $SO(1)$
instanton, whose only neutral zero-modes are $x^{\mu}$ and
$\theta_{\alpha}$, associated to the superspace coordinates.

Actually, one can define four different types of ${\rm E}1_3$
instantons, corresponding to the four inequivalent choices of
Chan-Paton factors
\begin{equation}
\begin{split}
&D_{h;o}^1=r_3\,,
\\
&D_{h;g}^1=r_3 \,,
\\
& D_{h;f}^1=r_3\,,
\\
& D_{h;h}^1=r_3\,,
\\
\end{split}\qquad
\begin{split}
&D_{h;o}^2=r_3\,,
\\
&D_{h;g}^2=r_3\,,
\\
&D_{h;f}^2=-r_3\,,
\\
&D_{h;h}^2=-r_3\,,
\\
\end{split}\qquad
\begin{split}
&D_{h;o}^3=r_3\,,
\\
&D_{h;g}^3=-r_3\,,
\\
&D_{h;f}^3=r_3\,,
\\
&D_{h;h}^3=-r_3\,,
\\
\end{split}\qquad
\begin{split}
&D_{h;o}^4=r_3\,,
\\
&D_{h;g}^4=-r_3\,,
\\
&D_{h;f}^4=-r_3\,,
\\
&D_{h;h}^4=r_3\,.
\\
\end{split}
\end{equation}
These are the gauge instantons for the four types of fractional D$5_3$ branes in the model in \cite{aadds}.
In the following we shall refer to them by ${\rm E}1_o$, ${\rm
E}1_g$, ${\rm E}1_f$ and ${\rm E}1_h$, where we have dropped the
index 3, since these are the only ones whose effects we shall study
in this paper.

Details on the partition function for E-branes can be found
in Appendix C, while table \ref{Multinst} summarises the structure
of the zero modes charged with respect to $G_{\rm CP}$, for the
simple case of a non-magnetised ${\rm E}1$ instanton.

\begin{table}[htdp]
\begin{center}
\begin{tabular}{|c|c|c|}
\hline
Instanton & Multiplicity & Representation\\[1ex]
\hline
\hline
${\rm E}1_o$&  $\frac{1}{4}\left(I^{ab}- S^{ab}_g I_1^{ab}- S^{ab}_f I_2^{ab}+ S^{ab}_h I_3^{ab}\right)$ & $(r_3,\bar p_b)$
\\ [1ex]
&  $\frac{1}{4}\left(I^{a\beta} - S^{\alpha \beta}_g I_1^{a\beta}+S^{\alpha \beta}_f I_2^{a\beta}-S^{\alpha \beta}_h I_3^{a\beta}\right)$ & $(r_3,\bar{q}_{\beta})$
\\ [1ex]
\hline
${\rm E}1_g$ &   $\frac{1}{4}\left(I^{ab}-S^{ab}_g I_1^{ab}+S^{ab}_f I_2^{ab}-S^{ab}_h I_3^{ab}\right)$ & $(r_3,\bar p_b)$
\\[1ex]
 ~& $\frac{1}{4}\left(I^{a\beta}-S^{\alpha \beta}_g I_1^{a\beta}-S^{\alpha \beta}_f I_2^{a\beta}+S^{\alpha \beta}_h I_3^{a\beta}\right)$ & $(r_3,\bar{q}_{\beta})$
 \\[1ex]
\hline
${\rm E}1_f$ &   $\frac{1}{4}\left(I^{ab}+S^{ab}_g I_1^{ab}-S^{ab}_f I_2^{ab}-S^{ab}_h I_3^{ab}\right)$ & $(r_3,\bar p_b)$
\\[1ex]
~& $\frac{1}{4}\left(I^{a\beta}+S^{\alpha \beta}_g I_1^{a\beta}+S^{\alpha \beta}_f I_2^{a\beta}+S^{\alpha \beta}_h I_3^{a\beta}\right)$ & $(r_3,\bar{q}_{\beta})$
\\[1ex]
\hline
${\rm E}1_h$ &   $\frac{1}{4}\left(I^{ab}+S^{ab}_g I_1^{ab}+S^{ab}_f I_2^{ab}+S^{ab}_h I_3^{ab}\right)$ & $(r_3,\bar p_b)$
\\[1ex]
~& $\frac{1}{4}\left(I^{a\beta}+S^{\alpha \beta}_g I_1^{a\beta}-S^{\alpha \beta}_f I_2^{a\beta}-S^{\alpha \beta}_h I_3^{a\beta}\right)$ & $(r_3,\bar{q}_{\beta})$
\\[1ex]
\hline
\end{tabular}
\end{center}
\caption{Charged zero modes for ${\rm E}1_3$ branes.  The index $a$
in the multiplicities corresponds to the E1$_3$ instantons, with
wrapping numbers $(m,n)  = \{ (1,0) \,,\,  (-1,0) \,,\, (0,1)\}$.}
\label{Multinst}
\end{table}

Finally, from the transverse channel amplitudes one can derive their gauge kinetic functions, that read
\begin{equation}
\begin{split}
&S_{{\rm E}1_o} = \sigma \left[T_3+4\alpha_1\, \left(- \sum_{l\in F_g} \epsilon_l^{(i),g}~ M^l_1+
\sum_{l\in F_f} \epsilon_l^{(i),f}~M^l_2\right)+4 \alpha_2 \, \sum_{l\in F_h} \epsilon_l^{(i),h}~M^l_3 \right] \ ,
\\
&S_{{\rm E}1_g} = \sigma \left[T_3+4\alpha_1 \, \left(- \sum_{l\in F_g} \epsilon_l^{(i),g}~M^l_1- \sum_{l\in F_f} \epsilon_l^{(i),f}~M^l_2\right)-4 \alpha_2\,  \sum_{l\in F_h} \epsilon_l^{(i),h}~M^l_3 \right] \ ,
\\
&S_{{\rm E}1_f} = \sigma \left[T_3+4\alpha_1 \, \left( \sum_{l\in F_g} \epsilon_l^{(i),g}~M^l_1+ \sum_{l\in F_f} \epsilon_l^{(i),f}~M^l_2\right)-4 \alpha_2 \, \sum_{l\in F_h} \epsilon_l^{(i),h}~M^l_3 \right] \ ,
\\
&S_{{\rm E}1_h} = \sigma \left[T_3+4\alpha_1 \, \left(\sum_{l\in F_g} \epsilon_l^{(i),g}~M^l_1- \sum_{l\in F_f} \epsilon_l^{(i),f}~M^l_2\right)+4 \alpha_2 \, \sum_{l\in F_h} \epsilon_l^{(i),h}~M^l_3 \right] \ ,
\\
\end{split}
\label{action}
\end{equation}
where $\sigma$ is a positive normalisation constant, and the
closed-string fields have been defined in the previous section.

\section{Explicit models}

In this section we shall analyse in detail some explicit orientifold configurations together with their non-perturbative corrections. For simplicity, we shall confine our attention to the case of equal magnetic fields on the brane stacks labelled by $a$ and $\alpha$. In the T-dual picture of D6 branes at angles, this would correspond to configurations where the  $a$ and $\alpha$ branes are rotated by the same angles, and thus, belong to the same equivalence class. This assumption will considerably simplify the twisted tadpole conditions (\ref{twisted}), but nevertheless will allow for some interesting physics. Therefore, in the following we shall suppress the index $\alpha$ and assume that $p_a=q_a$.
One such stack of magnetised D9-branes will then yield a $U(p_a)\times U(q_a)$ gauge group. In this Section, we shall use the index structure $^{(a)}_1$ to refer to a $U(p_a)$ factor and the index structure $^{(a)}_2$ to refer to a $U(q_a)$ factor.
The parametrisation of the Chan-Paton factors is then
\begin{equation}
\begin{split}
&N_{a,o}=p_a+q_a+\bar p_a+\bar q_a \,,
\\
&N_{a,f}=i(p_a-q_a-\bar p_a+\bar q_a)\,,
\\
\end{split}
\qquad
\begin{split}
&N_{a,g}=i(p_a+q_a-\bar p_a-\bar q_a) \,,
\\
&N_{a,h}=p_a-q_a+\bar p_a-\bar q_a \,.
\\
\end{split}
\end{equation}
The annulus and M\"obius strip partition functions can be easily derived from those in Appendix B, so to obtain the massless spectrum listed in table \ref{examples}. Notice that in the multiplicity of the bi-fundamental representation $(p_a , q_a)$ we have explicitly counted the number of brane intersections located at orbifold fixed points, since $p_a$ and $q_a$ branes have homologous wrapping numbers.

\begin{table}[htdp]
\begin{center}
\begin{tabular}{|c|c|c|}
\hline
Multiplicity & Representation & Relevant Indices
\\
\hline
\hline
$\frac{1}{8}(I^{aa'}+I^{aO}-4I_1^{aa'}-4I_2^{aa'}+4I_3^{aa'})$ & $\left(\frac{p_a(p_a-1)}{2},1\right)+\left(1,\frac{q_a(q_a-1)}{2}\right)$ & $\forall a$
\\[1ex]
$\frac{1}{8}(I^{aa'}-I^{aO}-4I_1^{aa'}-4I_2^{aa'}+4I_3^{aa'})$ & $\left(\frac{p_a(p_a+1)}{2},1\right)+\left(1,\frac{q_a(q_a+1)}{2}\right)$ & $\forall a$
\\[1ex]
\hline
$\frac{1}{4}(I^{aa'}-4I_1^{aa'}+4I_2^{aa'}-4I_3^{aa'})$& $(p_a,q_a)$ & $\forall a$
\\[1ex]
$\frac{1}{4}(I^{ab}+S_g^{ab}I_1^{ab}-S_f^{ab}I_2^{ab}-  S_h^{ab}I_3^{ab})$& $(p_a,\bar{q}_b)$ & $a \ne b$
\\[1ex]
$\frac{1}{4}(I^{ab'}-S_g^{ab}I_1^{ab'}+S_f^{ab}I_2^{ab'}-S_h^{ab}I_3^{ab'})$& $(p_a,q_b)$  & $a \ne b$
\\[1ex]
\hline
$\frac{1}{4}(I^{ab'}-S_g^{ab}I_1^{ab'}-S_f^{ab}I_2^{ab'}+S_h^{ab}I_3^{ab'})$ & $(p_a,p_b)+(q_a,q_b)$ & $a < b$
\\[1ex]
$\frac{1}{4}(I^{ab}+S_g^{ab}I_1^{ab}+S_f^{ab}I_2^{ab}+S_h^{ab}I_3^{ab})$& $(p_a,\bar{p}_b)+(q_a,\bar{q}_b)$ & $a < b$
\\[1ex]
\hline
$1$ & $(p_a,\bar q_a)+(\bar p_a,q_a)$ & $\forall a$
\\[1ex]
\hline
\end{tabular}
\end{center}
\caption{Representations and multiplicities of charged chiral superfields on a $T^6 /\mathbb{Z}_2 \times \mathbb{Z}_2$ orbifold with discrete torsion, in the presence of equal magnetic fields on ${\rm D}9^{(a)}$ and ${\rm D}9^{(\alpha)}$ branes.}
\label{examples}
\end{table}

\subsection{Fractional branes and instanton breaking of conformal  invariance}

The first model we consider was first built in \cite{cristina} and, has only two stacks of branes with the following wrapping
numbers
\begin{equation}
\begin{split}
&(m_i^{(1)},n_i^{(1)}) \ = \{ (1,1) \,,\, (1,1) \,,\, (-1,1) \} \,,
\\
&(m_i^{(2)},n_i^{(2)}) \ = \{ (-1,1) \,,\, (-1,1) \,,\, (1,1)\} \,.
\\
\end{split}
\label{intersectionconformal}
\end{equation}
We note that although in a toroidal setup these two stacks would be merely the images of each other.  In the present setup, their different twisted charges, crucial for the cancelation of the twisted tadpoles in eq.~(\ref{twisted}),  make them physically inequivalent.
The gauge group is therefore $U(4)^2\times U(4)^2$ and the massless spectrum consists
of chiral multiplets in the representations
\begin{equation}
\begin{split}
&(\bar 4,4,1,1)_{(-1,1,0,0)}+(4,\bar 4,1,1)_{(1,-1,0,0)} + (1,1,\bar 4,4)_{(0,0,-1,1)}+(1,1,4,\bar 4)_{(0,0,1,-1)}
\\
& + 8\times[(\bar 6,1,1,1)_{(-2,0,0,0)}+(1,\bar 6,1,1)_{(0,-2,0,0)}+(1,1,6,1)_{(0,0,2,0)}+(1,1,1,6)_{(0,0,0,2)}] \,,
\\
& + (4,1,4,1)+(\bar 4 , 1, \bar 4 , 1) + (4,1,1,4) + (\bar 4 , 1 ,1 , \bar 4) \\
&+ (1,4,4,1)
+ (1,\bar 4 , \bar 4 , 1) + (1,4,1,4) + (1 ,\bar 4, 1 , \bar 4)
\\
\end{split}
\end{equation}
where the subscript clearly denotes the $U(1)^4$ charges. Actually, all Abelian factors are anomalous, and therefore the low-energy group is $SU(4)^4$.

The contribution of the anomaly encoded in the matrix
\begin{equation}
C_{ab}=\frac{32}{\pi^2} \, \left(
  \begin{array}{cccc}
    -1 & 0 & 0 & 0 \\
    0 & -1 & 0 & 0 \\
    0 & 0 & 1 & 0 \\
    0 & 0 & 0 & 1 \\
  \end{array}
\right) \,,
\label{anommatrix11}
\end{equation}
must then be cancelled by the gauge variations of the gauge kinetic functions
\begin{equation}
\begin{split}
&f_1^{(1)} \, = \, S+T_1+T_2-T_3+\alpha_1\, (M_1+M_2) + \alpha_2 \, M_3 \,,
\\
&f_2^{(1)} \, = \, S+T_1+T_2-T_3+\alpha_1\, (M_1-M_2) - \alpha_2 \, M_3 \,,
\\
&f_1^{(2)} \, = \, S+T_1+T_2-T_3 - \alpha_1 \, (M_1+M_2) + \alpha_2 \, M_3 \,,
\\
&f_2^{(2)} \, = \, S+T_1+T_2-T_3-\alpha_1\, (M_1-M_2) - \alpha_2 \, M_3 \,.
\\
\end{split}
\end{equation}
Using the following expression for the gauge variations from eq.(\ref{variation4})
\begin{equation}
\begin{split}
&\delta S =
4\, \lambda_S\,\left[-\varLambda^{(1)}_1-\varLambda^{(1)}_2+\varLambda^{(2)}_1+\varLambda^{(2)}_2\right] \,,
\\
&\delta T_1=\delta T_2=-\delta T_3=
4\, \lambda\, \left[\varLambda^{(1)}_1+\varLambda^{(1)}_2-\varLambda^{(2)}_1-\varLambda^{(2)}_2\right] \,,
\\
&\delta M_1=4\, \lambda_1\, \alpha_1\,
\left[\varLambda^{(1)}_1+\varLambda^{(1)}_2+\varLambda^{(2)}_1+\varLambda^{(2)}_2\right] \,,
\\
&\delta M_2=4\, \lambda_2\, \alpha_1\,
\left[\varLambda^{(1)}_1-\varLambda^{(1)}_2+\varLambda^{(2)}_1-\varLambda^{(2)}_2\right]
\,,
\\
&\delta M_3=4\,\lambda_3\,\alpha_2\, \left[-\varLambda^{(1)}_1+\varLambda^{(1)}_2+\varLambda^{(2)}_1-\varLambda^{(2)}_2\right]\,,
\\
\end{split}
\end{equation}
and taking into account the fact that we work with non-canonically normalised $U(1)$ generators, one obtains the following expressions for $C_{ab}$
\begin{equation}
\begin{split}
&C_{11}=C_{22}=-C_{33}=-C_{44}=32\, \left(-\lambda_S+3\,\lambda+\alpha_1^2\,\lambda_1+\alpha_1^2\,\lambda_2-\alpha_2^2\,\lambda_3\right) \,,
\\
&C_{12}=C_{21}=-C_{34}=-C_{43}=32\, \left(-\lambda_S+3\,\lambda+\alpha_1^2\,\lambda_1-\alpha_1^2\,\lambda_2+\alpha_2^2\,\lambda_3\right) \,,
\\
&C_{13}=C_{24}=-C_{31}=-C_{42}=32\,\left(\lambda_S-3\,\lambda+\alpha_1^2\,\lambda_1+\alpha_1^2\,\lambda_2+\alpha_2^2\,\lambda_3\right) \,,
\\
&C_{14}=C_{23}=-C_{32}=-C_{41}=32\,\left(\lambda_S-3\,\lambda+\alpha_1^2\,\lambda_1-\alpha_1^2\,\lambda_2-\alpha_2^2\,\lambda_3\right) \,.
\\
\end{split}
\label{anommatrix1}
\end{equation}
Requiring that eq. (\ref{anommatrix1}) matches the anomalous contributions encoded in eq. (\ref{anommatrix11}) one finds
\begin{equation}
-\lambda_S+3\,\lambda=\alpha_1^2\,\lambda_1=\alpha_1^2\,\lambda_2=-\alpha_2^2\,\lambda_3=-
\frac{1}{4\pi^2}
\,.
\label{sys1}
\end{equation}

We can now turn to the analysis of instantonic effects induced by the ${\rm E}1_3$ brane.
We denote the chiral fields in the antisymmetric ($6$ and $\bar 6$) representations by $A^1_{ij}$, $A^2_{ij}$, $A^3_{ij}$ and $A^4_{ij}$, respectively. The instantonic sector is labelled by the vector $(k_1,k_2,k_3,k_4)$, where each $k_i$ is the instanton number associated to the $i$-th unitary gauge group. As already stressed in the previous section, we are essentially interested to the effect of (single) $SO(1)$ instantons, and therefore we assume $k_i =0,1$. The zero-mode structure of these E1 instantons is
derived in Appendix C, and is then listed in table \ref{inst1table}

\begin{table}[htdp]
\begin{center}
\begin{tabular}{|c|c|c|c|}
\hline
Instanton & $(k_1,k_2,k_3,k_4)$ & Representation & Zero mode
\\ [1ex]
\hline
\hline
${\rm E}1_o$ & $(1,0,0,0)$ & $(1,4,1,1)$ & $\eta^{o}_i$
\\[1ex]
${\rm E}1_g$ & $(0,1,0,0)$ & $(4,1,1,1)$ & $\eta^{g}_i$
\\[1ex]
${\rm E}1_f$ & $(0,0,1,0)$ & $(1,1,\bar 4,1)$ & $\eta^{f}_i$
\\[1ex]
${\rm E}1_h$ & $(0,0,0,1)$ & $(1,1,1,\bar 4)$ & $\eta^{h}_i$
\\[1ex]
\hline
\end{tabular}
\end{center}
\caption{Charged zero-mode structure for $SO(1)$ E1 instantons for models with only fractional branes. The index $i$ runs over the (anti)fundamental representation of $U(4)$.}
\label{inst1table}
\end{table}

Let us analyse the case of a single ${\rm E}1_o$ instanton in detail. The gauge-invariant instantonic action including both neutral and charged zero modes is
\begin{equation}
S_{\rm inst} = S_{{\rm E}1_o} + \sum_{i,j=1}^4\eta^{o}_iA^2_{ij}\eta^{o}_j\,,
\end{equation}
where $S_{{\rm E}1_o} $ was derived in Section 3 and is given explicitly in eq. (\ref{action}). Upon integration over the charged instantonic zero-modes
\begin{equation}
\int\prod_{i=1}^4d\eta_i^{o} \, e^{-S_{\rm inst}} \,,
\end{equation}
a non-perturbative correction to the superpotential is generated
\begin{equation}
\mathcal{W}_{\rm non-pert} = e^{- S_{{\rm E}1_o}}
\sum_{i,j,k,l=1}^4\epsilon_{ijkl}A_{ij}^2A_{kl}^2 \,,
\label{sys4}
\end{equation}
that corresponds to a mass term for $A_{ij}^2$.

Similar results, clearly hold when the other instantonic contributions are taken into account. In particular,  a single ${\rm E}1_g$ instanton gives mass to $A^1_{ij}$, a single ${\rm E}1_f$ instanton gives mass to $A^3_{ij}$, and finally
a single ${\rm E}1_h$ instanton gives mass to $A^4_{ij}$.

Demanding that the new coupling in the superpotential (\ref{sys4}) be gauge invariant, actually fixes the overall normalisation of the instantonic action and of the gauge variation for the K\"ahler modulus $T_3$. In fact, from eqs. (\ref{variation4}) and (\ref{action}) one can easily determine the gauge variation of $S_{{\rm E}1_o}$
\begin{equation}
\delta S_{{\rm E}1_o} =
4 \sigma  \left[- \left( \lambda + \frac{1}{\pi^2} \right)\, \varLambda^{(1)}_1-\left(\lambda-\frac{3}{\pi^2} \right)\, \varLambda^{(1)}_2
 + \left( \lambda+\frac{1}{\pi^2} \right) \varLambda^{(2)}_1+\left( \lambda+\frac{1}{\pi^2}\right)\, \varLambda^{(2)}_2\right] \,,
\end{equation}
and therefore one must choose
\begin{equation}
\lambda  = - \frac{1}{\pi^2} = - \frac{1}{4\sigma}\,,
\label{sys2}
\end{equation}
so that $\delta S_{{\rm E}1_o} =4\, \Lambda^{(1)}_2$, and $e^{-S_{{\rm E}1_o}}AA$ is indeed gauge invariant. Moreover,
eqs. (\ref{sys1}) and (\ref{sys2}) also fix
\begin{equation}
\lambda_S = - 11/4\pi^2\,.
\label{sys3}
\end{equation}
Notice that this is consistent with the gauge invariance of the non-perturbative contributions from the other instantons, since
\begin{equation}
\begin{split}
&\delta S_{{\rm E}1_g} =
4 \sigma  \left[- \left( \lambda - \frac{3}{\pi^2} \right)\, \varLambda^{(1)}_1-\left(\lambda+\frac{1}{\pi^2} \right)\, \varLambda^{(1)}_2
 + \left( \lambda+\frac{1}{\pi^2} \right) \varLambda^{(2)}_1+\left( \lambda+\frac{1}{\pi^2}\right)\, \varLambda^{(2)}_2\right] \,,
\\
&\delta S_{{\rm E}1_f} =
4 \sigma  \left[- \left( \lambda + \frac{1}{\pi^2} \right)\, \varLambda^{(1)}_1-\left(\lambda+\frac{1}{\pi^2} \right)\, \varLambda^{(1)}_2
 + \left( \lambda-\frac{3}{\pi^2} \right) \varLambda^{(2)}_1+\left( \lambda+\frac{1}{\pi^2}\right)\, \varLambda^{(2)}_2\right] \,,
\\
&\delta S_{{\rm E}1_h} =
4 \sigma  \left[- \left( \lambda + \frac{1}{\pi^2} \right)\, \varLambda^{(1)}_1-\left(\lambda+\frac{1}{\pi^2} \right)\, \varLambda^{(1)}_2
 + \left( \lambda+\frac{1}{\pi^2} \right) \varLambda^{(2)}_1+\left( \lambda-\frac{3}{\pi^2}\right)\, \varLambda^{(2)}_2\right] \,,
\\
\end{split}
\end{equation}
and therefore the same choice (\ref{sys2}) yields gauge-invariant mass terms for the corresponding open-string fields in the antisymmetric representations.

It is particularly interesting to study a minor deformation of this
model, where the second pair of D9 branes is endowed with suitable
Wilson lines. In order to still solve the twisted tadpole conditions,
Wilson lines can be turned on only along the first $T^2$ that is
unaffected by the $g$ twist. More precisely, the Wilson lines are
T-dual to a motion placing them in the other fixed points of the $f$ and $h$
operations The resulting spectrum divides into two
non-interacting sectors. Both sets of branes have effective gauge group $SU(4)^2$, chiral
supermultiplets $\varPhi$ and $\tilde \varPhi$ in the bi-fundamental
representations $(\bar 4, 4)$ and $(4,\bar 4)$ and eight fields $A^1$ and $A^2$ in the anti-symmetric representations $(\bar 6,1)$ and $(1,\bar 6)$.
The beta functions for the non-Abelian gauge couplings are vanishing
at one loop, and there are no renormalisable Yukawa couplings. As a
result, the theory is conformal at one-loop, also conformal invariance
is broken at  higher loop order. However, it is tempting to speculate
about the existence of an IR conformal fixed point, following the
arguments of \cite{ls}. Indeed, the all-loop beta functions for gauge couplings are proportional to
\begin{equation}
\beta_g = 3 C_2 (G) - \sum_R T (R) + \sum_R T (R) \gamma (\varPhi_R) \,,
\end{equation}
so that in the present case, assuming that the eight fields in the anti-symmetric representation all share the same anomalous dimension, the two gauge couplings $g_1$ and $g_2$ run according to
\begin{equation}
\beta_1 (g_1,g_2) \, = \, 2 (\gamma_{\varPhi} + \gamma_{\tilde \varPhi} + 4 \gamma_{A^1}) \, , \quad
\beta_2 (g_1,g_2) \ =  \ 2 (\gamma_{\varPhi} + \gamma_{\tilde \varPhi} + 4 \gamma_{A^2})\,.
\label{conformal1}
\end{equation}
Since $\gamma_{\varPhi} = \gamma_{\tilde \varPhi}$, conformal invariance persists at all orders if
\begin{equation}
\begin{split}
&\gamma_{\varPhi} + 2 \gamma_{A^1} = 0 \,,
\\
&\gamma_{\varPhi} + 2 \gamma_{A^2}=0 \,.
\\
\end{split}
\end{equation}
This is a system of two equations in two unknowns, $g_1$ and $g_2$, that might in principle admit a solution. Clearly, the origin
$g_1=g_2=0$ is a solution, while if $(g_1^0 ,g_2^0)$ is a discrete solution than so is $(g_2^0 ,g_1^0)$, because of the relation
$ \gamma_{A^1} (g_1,g_2) = \gamma_{A^2} (g_2,g_1)$. These discrete solutions would then correspond to IR conformal fixed points.
Moreover, since the two eqs. (\ref{conformal1}) are independent, one can conclude that the theory has no conformal fixed line.

We have just shown, however, that instantons generate a mass term for the fields in the antisymmetric representations. Although, these terms clearly break conformal invariance at the perturbative level, non-perturbatively this is not so obvious, since a quadratic coupling
$m AA$ in the superpotential might still be conformal provided its anomalous dimension vanishes,
\begin{equation}
\beta_m \ \sim \ - 1 \ + \ \frac{1}{2}\ \gamma_{A^{1,2}} \ = \ 0 \,.
\label{conformal2}
\end{equation}
However, the system of equations (\ref{conformal1}) and (\ref{conformal2}) is overconstrained
and {\it a priori} has no solution, and we can conclude that conformal invariance is actually broken by non-perturbative effects at a
hierarchically small energy scale. This model is actually a supersymmetric variant of the Frampton and Vafa approach to the
 hierarchy problem \cite{fv}, where quantum corrections to the electroweak scale are protected by conformal symmetry, that is
 broken at a scale of the order of a TeV. In our supersymmetric variant, the emergence of a non-perturbative mass term decouples some
 states in the IR, and thus yields an effective low-energy theory with broken conformal symmetry. Needless to say, it would be
 interesting to provide concrete non-supersymmetric models of this kind with the low-energy gauge group and spectrum close to that
 of the Standard Model.

\subsection{Models with fractional and bulk branes}

The second model we want to describe, is a modification of the previous one
where some of the fractional branes are replaced by regular (bulk) branes away from the fixed points, but still endowed with
non-trivial magnetic fluxes preserving $\mathcal{N}=1$ supersymmetry. The gauge group associated with these bulk branes is $U(n)$ and the additional light excitations are listed in table \ref{bulkbranes}

\begin{table}[htdp]
\begin{center}
\begin{tabular}{|c|c|}
\hline
 Multiplicity & Representation
 \\[1ex]
\hline
\hline
 3 & $ n\bar n$
 \\ [1ex]
 $2(I^{aa'} + \frac{1}{4}I^{aO})$&$ \frac{n(n-1)}{2}$
 \\[1ex]
 $2(I^{aa'}- \frac{1}{4} I^{aO}) $& $\frac{n(n+1)}{2}$
 \\[1ex]
 \hline
 $I^{ab}$  &$(n,\bar{p}_b)+(n,\bar{q}_b)$
 \\[1ex]
 $I^{ab'}$ &$(n,p_b)+(n,q_b)$
 \\[1ex]
\hline
\end{tabular}
\end{center}
\caption{Representations and multiplicities of chiral superfields on bulk branes $a$. The intersection numbers have
the same definition as in eqs. (\ref{intersections}) and (\ref{ointersections}).}
\label{bulkbranes}
\end{table}

A possible solution to the tadpole (and supersymmetry) conditions, consists of two pairs of fractional branes with intersection numbers
(\ref{intersectionconformal}) together with two coincident bulk branes with
\begin{equation}
(m_i^{(3)},n_i^{(3)})= \{ (1,1)\,,\, (1,1)\,,\, (-1,1)\}\,.
\end{equation}
The gauge group is $U(2)^2\times U(2)^2\times U(2)_{\rm bulk}$ and the charged massless spectrum
consists of chiral multiplets in the representations listed in table \ref{instbulkspectr}, where names for the relevant fields are explicitly given.

\begin{table}[htdp]
\begin{center}
\begin{tabular}{|c|c|c|}
\hline
 Multiplicity & Representation & Field
 \\[.5ex]
\hline
\hline
1&$(\bar2,2,1,1,1)_{(-1,1,0,0,0)}$ &
\\[.3ex]
1&$(2,\bar2,1,1,1)_{(1,-1,0,0,0)}$&
\\[.3ex]
1&$(1,1,\bar2,2,1)_{(0,0,-1,1,0)}$ &
\\[.3ex]
1&$(1,1,2,\bar2,1)_{(0,0,1,-1,0)}$ &
\\[.3ex]
3&$(1,1,1,1,4)_{(0,0,0,0,0)}$&
\\[.3ex]
8& $(\bar1,1,1,1,1)_{(-2,0,0,0,0)}$ & $A^1$
\\[.3ex]
8&$(1,\bar1,1,1,1)_{(0,-2,0,0,0)}$ & $A^2$
\\[.3ex]
8&$(1,1,1,1,1)_{(0,0,2,0,0)}$ & $A^3$
\\[.3ex]
8&$(1,1,1,1,1)_{(0,0,0,2,0)}$ & $A^4$
\\[.3ex]
32& $(1,1,1,1,\bar1)_{(0,0,0,0,-2)}$&$A^5$
\\[.3ex]
8&$(1,1,2,1,\bar2)_{(0,0,1,0,-1)}$&$\varPhi_{3\bar5}$
\\[.3ex]
8&$(1,1,1,2,\bar2)_{(0,0,0,1,-1)}$&$\varPhi_{4\bar5}$
\\[.3ex]
8&$(\bar2,1,1,1,\bar2)_{(-1,0,0,0,-1)}$&$\varPhi_{\bar1\bar5}$
\\[.3ex]
8&$(1,\bar2,1,1,\bar2)_{(0,-1,0,0,-1)}$&$\varPhi_{\bar2\bar5}$
\\[.3ex]
\hline
\end{tabular}
\end{center}
\caption{Massless spectrum charged with respect to the $U(2)^2 \times U(2)^2 \times U(2)_{\rm bulk}$ gauge group.}
\label{instbulkspectr}
\end{table}

The anomaly matrix is now
\begin{equation}
C_{ab}=\frac{8}{\pi^2}\, \left(
  \begin{array}{ccccc}
    -1 & 0 & 0 & 0 & -1 \\
    0 & -1 & 0 & 0 & -1 \\
    0 & 0 & 1 & 0 & -1 \\
    0 & 0 & 0 & 1 & -1 \\
    -1 & -1 & 1 & 1 & -4 \\
  \end{array}
\right) \,,
\end{equation}
and a generalised Green-Schwarz mechanism demands that the closed-string fields transform non-linearly under the gauge transformations as,
\begin{equation}
\begin{split}
&\delta S =
2\, \lambda_S\, \left[-\varLambda^{(1)}_1-\varLambda^{(1)}_2+\varLambda^{(2)}_1+\varLambda^{(2)}_2-\gamma \, \varLambda^{(3)}\right] \,,
\\
&\delta T_1 = \delta T_2=-\delta T_3 =
2\, \lambda\, \left[\varLambda^{(1)}_1+\varLambda^{(1)}_2-\varLambda^{(2)}_1-\varLambda^{(2)}_2
+\gamma\, \varLambda^{(3)}\right] \,,
\\
&\delta M_1 = 2\,\lambda_1\, \alpha_1\,
\left[\varLambda^{(1)}_1+\varLambda^{(1)}_2+\varLambda^{(2)}_1+\varLambda^{(2)}_2\right]\,,
\\
&\delta M_2=2\, \lambda_2 \,\alpha_1\,
\left[\varLambda^{(1)}_1-\varLambda^{(1)}_2+\varLambda^{(2)}_1-\varLambda^{(2)}_2\right]\,,
\\
&\delta M_3=2\,\lambda_3\, \alpha_2 \,\left[-\varLambda^{(1)}_1+\varLambda^{(1)}_2+\varLambda^{(2)}_1-\varLambda^{(2)}_2\right]\,,
\\
\end{split}
\end{equation}
so that the variation of the gauge kinetic functions
\begin{equation}
\begin{split}
&f^{(1)}_1 \, = \, S+T_1+T_2-T_3+\alpha_1 \, (M_1+M_2)+\alpha_2 \,M_3 \,,
\\
&f^{(1)}_2\, = \, S+T_1+T_2-T_3+\alpha_1 \, (M_1-M_2)-\alpha_2\, M_3 \,,
\\
&f^{(2)}_1 \, = \, S+T_1+T_2-T_3-\alpha_1 \, (M_1+M_2)+ \alpha_2 \, M_3 \,,
\\
&f^{(2)}_2 \, = \, S+T_1+T_2-T_3-\alpha_1 \, (M_1-M_2)-\alpha_2 \, M_3 \,,
\\
&f^{(3)} \, = \, \gamma \, (S+T_1+T_2-T_3) \,,
\\
\end{split}
\end{equation}
can cancel the anomaly.
Taking into account that the $U(1)$ generators are not canonically normalised we arrive at the following set of equations
\begin{equation}
\begin{split}
&C_{11}=C_{22}=-C_{33}=-C_{44}=8 \left(-\lambda_S+3\lambda+\alpha_1^2\lambda_1+\alpha_1^2\lambda_2-\alpha_2^2\lambda_3
\right) \,,
\\
&C_{12}=C_{21}=-C_{34}=-C_{43}=8 \left(-\lambda_S+3\lambda+\alpha_1^2\lambda_1-\alpha_1^2\lambda_2+\alpha_2^2\lambda_3\right) \,,
\\
&C_{13}=C_{24}=-C_{31}=-C_{42}=8 \left(\lambda_S-3\lambda+\alpha_1^2\lambda_1+\alpha_1^2\lambda_2+\alpha_2^2\lambda_3\right) \,,
\\
&C_{14}=C_{23}=-C_{32}=-C_{41}=8
\left(\lambda_S-3\lambda+\alpha_1^2\lambda_1-\alpha_1^2\lambda_2-\alpha_2^2\lambda_3\right) \,,
\\
&C_{15}=C_{25}=C_{35}=C_{45}=C_{51}=C_{52}=-C_{53}=-C_{54}= \gamma^{-1}\, C_{55}=8\gamma(-\lambda_S+3\lambda)
\,.
\\
\end{split}
\end{equation}
This system has (of course) the same solution eq. (\ref{sys1}) as the previous model. Notice
that $\gamma=4$ is uniquely singled out from anomaly cancelation conditions, in agreement with the
fact that the bulk branes are actually quadruplets with respect to the orientifold group.

We can now turn to the analysis of non-perturbative effects induced by the ${\rm E}1_3$ branes. We shall restrict our attention only to rigid instantons associated to fractional branes, that we label by the vector $(k_1 , k_2 , k_3 , k_4)$.
As in the previous case, we consider single instanton contributions, corresponding to branes with $SO(1)$ Chan-Paton group, so that $k_i =0,1$.  The zero-mode structure of these ${\rm E}1$ instantons is listed in table \ref{bulkinstzero}.

\begin{table}[htdp]
\begin{center}
\begin{tabular}{|c|c|c|c|}
\hline
Instanton & $(k_1,k_2,k_3,k_4)$ & Representation & Zero mode
\\[1ex]
\hline
\hline
${\rm E}1_o$ & $(1,0,0,0)$ & $(1,2,1,1,1),\ (1,1,1,1,2)$ & $\eta^{o}_i$,\ $\eta^{b}_i$
\\[1ex]
${\rm E}1_g$ & $(0,1,0,0)$ & $(2,1,1,1,1),\ (1,1,1,1,2)$ & $\eta^{g}_i$,\ $\eta^{b}_i$
\\[1ex]
${\rm E}1_f$ & $(0,0,1,0)$ & $(1,1,\bar 2,1,1),\ (1,1,1,1,2)$ & $\eta^{f}_i$,\ $\eta^{b}_i$
\\[1ex]
${\rm E}1_h$ & $(0,0,0,1)$ & $(1,1,1,\bar 2,1),\ (1,1,1,1,2)$ & $\eta^{h}_i$,\ $\eta^{b}_i$
\\[1ex]
\hline
\end{tabular}
\end{center}
\caption{Charged zero-mode structure for $SO(1)$ E1 instantons for models with fractional and bulk branes. The index $i$ refers to the (anti)fundamental representation of $U(2)$.}
\label{bulkinstzero}
\end{table}

Let us analyse the case of a single ${\rm E}1_o$ instanton in detail. The gauge-invariant instanton action including both neutral and charged zero modes is now
\begin{equation}
S_{\rm inst} = S_{{\rm E}1_o} + \sum_{i,j=1}^2 \left( \eta^{o}_i \, A^2_{ij}\, \eta^{o}_j + \eta^{b}_i \, A^5_{ij} \, \eta^{b}_j
+ \eta^{o}_i \, (\varPhi_{\bar2\bar5})_{ij} \, \eta^{b}_j \right) \,.
\end{equation}
Upon integration over the charged zero modes $\eta^o_i$ and $\eta^b_i$, one gets the following non-perturbative contribution to the superpotential
\begin{equation}
\mathcal{W}_{\rm non-pert} = e^{- S_{{\rm E}1_o}} \,
\sum_{i,j,k,l=1}^2\epsilon_{ij}\epsilon_{kl}\left[A_{ij}^2A_{kl}^5
-\frac{1}{2}(\varPhi_{\bar2\bar5})_{ik}(\varPhi_{\bar2\bar5})_{jl}\right]
\,.
\end{equation}

Similar results are obtained when considering the other instantons. For instance, an ${\rm E}1_g$ instanton induces mass terms of the form $A^1_{ij}A_{kl}^5-\frac{1}{2}(\varPhi_{\bar1\bar5})_{ik}(\varPhi_{\bar1\bar5})_{jl}$, an ${\rm E}1_f$ instanton induces mass terms of the form $A^3_{ij}A_{kl}^5-\frac{1}{2}(\varPhi_{3\bar5})_{ik}(\varPhi_{3\bar5})_{jl}$, and finally
an ${\rm E}1_h$ instanton induces mass terms of the form $A^4_{ij}A_{kl}^5-\frac{1}{2}(\varPhi_{4\bar5})_{ik}(\varPhi_{4\bar5})_{jl}$. Equations (\ref{variation4}), (\ref{action}),  (\ref{sys1}), (\ref{sys2}) and (\ref{sys3}) determine the gauge variations of the instanton actions
\begin{equation}
\begin{split}
&\delta S_{{\rm E}1_o} = 2\varLambda^{(1)}_2+2\varLambda^{(3)} \,,
\\
&\delta S_{{\rm E}1_f} = -2\varLambda^{(2)}_1+2\varLambda^{(3)} \,,
\\
\end{split}
\qquad
\begin{split}
&\delta S_{{\rm E}1_g} = 2\varLambda^{(1)}_1+2\varLambda^{(3)} \,,
\\
&\delta S_{{\rm E}1_h}=-2\varLambda^{(2)}_2+2\varLambda^{(3)}\,,
\\
\end{split}
\end{equation}
which imply, as expected, that the operators $e^{-S_{{\rm E}1}}(A^iA^5+\varPhi\varPhi)$ are gauge invariant.

\subsection{Models with linear terms in the superpotential}

Linear terms in the superpotential are necessary ingredients in F-term supersymmetry breaking, {\it \`a la}
O'Raifeartaigh for rigid supersymmetry or {\it \`a la} Polonyi in supergravity. Moreover, they have been recently employed to realise gauge mediation supersymmetry breaking within string theory \cite{mirjam}, and to achieve moduli stabilisation \cite{dmprt}.
Motivated by these applications, we finally study vacuum configurations where the magnetised branes intersect the ${\rm E}1_3$ instantons, while the remaining unmagnetised branes are parallel to them along one internal torus. Suitable Wilson lines can then make massive the charged zero-modes stretched between the E$1_3$ and unmagnetised branes, while linear terms in the superpotential are generated thanks to the presence of chiral fields in the antisymmetric representation of $U(2)$ fractional branes, coupling to the charged zero-modes stretched between the E$1_3$ and the fractional magnetised branes.

To be more explicit, the models include two stacks of magnetised D9 branes, one stack of non-magnetised D9 branes and a last stack of non-magnetised D5 branes. Their wrapping numbers are, respectively,
\begin{equation}
\begin{split}
&(m_i^{(1)},n_i^{(1)})= \{(2,1)\,,\, (1,1) \,,\, (-1,1) \} \,,
\\
&(m_i^{(2)},n_i^{(2)})= \{ (-2,1) \,,\, (-1,1) \,,\,(1,1) \} \,,
\\
&(m_i^{(3)},n_i^{(3)})= \{ (0,1) \,,\, (0,1) \,,\, (0,1)  \} \,,
\\
&(m_i^{(4)},n_i^{(4)})= \{ (0,1) \,,\, (1,0) \,,\, (-1,0)  \} \,.
\\
\end{split}
\end{equation}
Tadpole conditions then select the Chan-Paton gauge group
\begin{equation}
G_{\rm CP} = U(2)^2\times U(2)^2 \times USp(4)^2 \times USp(4)^2\,,
\label{cpgroup}
\end{equation}
where pairs of unitary groups live on each fractional brane, while the symplectic ones originate from non-magnetised D9 and D5 branes that, as anticipated, are displaced in the bulk. Charged matter corresponds to chiral supermultiplets in bi-fundamental and (anti)symmetric representations, with given multiplicities. Aside from the open strings stretched between fractional branes whose excitations are listed in table \ref{table1}, the model also includes open strings stretched between non magnetised (D9 and/or D5) branes and together with open strings stretched between non magnetised branes and fractional (magnetised) ones. Their generic massless excitations are listed in tables \ref{finalex1} and \ref{finalex2}, respectively.

\begin{table}[htdp]
\begin{center}
\begin{tabular}{|c|c|}
\hline
Multiplicity & Representation
\\[1ex]
\hline
\hline
1 & $\frac{n_1(n_1-1)}{2}$
\\[1ex]
1 & $\frac{n_2(n_2-1)}{2}$
\\[1ex]
2 & $(n_1,n_2)$
\\[1ex]
\hline
1 & $\frac{d_1(d_1-1)}{2}$
\\[1ex]
1 & $\frac{d_2(d_2-1)}{2}$
\\[1ex]
2 & $(d_1,d_2)$
\\[1ex]
\hline
1 & $(n_1,d_1)$
\\[1ex]
1 & $(n_1,d_2)$
\\[1ex]
1 & $(n_2,d_1)$
\\[1ex]
1 & $(n_2,d_2)$
\\[1ex]
\hline
\end{tabular}
\end{center}
\caption{Massless spectrum from open strings stretched between non magnetised D9 ($n_i$) and D5 ($d_i$) branes.}
\label{finalex1}
\end{table}

\begin{table}[htdp]
\begin{center}
\begin{tabular}{|c|c|}
\hline
Multiplicity & Representation
\\[1ex]
\hline
\hline
$\frac{1}{2}(I^{ab}+2I_3^{ab})$ & $(p_a,n_1)+(q_a,n_2)$
\\[1ex]
$\frac{1}{2}(I^{ab}-2I_3^{ab})$ & $(p_a,n_2)+(q_a,n_1)$
\\[1ex]
$\frac{1}{2}(I^{ab}+2I_2^{ab})$ & $(p_a,d_1)+(q_a,d_2)$
\\[1ex]
$\frac{1}{2}(I^{ab}-2I_2^{ab})$ & $(p_a,d_2)+(q_a,d_1)$
\\[1ex]
\hline
\end{tabular}
\end{center}
\caption{Massless spectrum from open strings stretched between non magnetised D9 ($n_i$) or D5 ($d_i$) branes and magnetised fractional D9 branes.}
\label{finalex2}
\end{table}

Inserting the explicit values of the wrapping numbers and the Chan-Paton gauge group eq. (\ref{cpgroup}) one finds the charged spectrum in table \ref{spectrum3}, where names for the relevant fields are explicitly given.

\begin{table}[htdp]
\begin{center}
\begin{tabular}{|c|c|c|}
\hline
Multiplicity & Representation & field
\\[.5ex]
\hline
\hline
1 & $(2,\bar2,1,1;1,1,1,1)$ & $\varPhi_{1\bar2}$
\\[.3ex]
1 & $(\bar2,2,1,1;1,1,1,1)$ & $\varPhi_{\bar12}$
\\[.3ex]
12 & $(\bar1,1,1,1;1,1,1,1)$ & $A^1$
\\[.3ex]
12 & $(1,\bar1,1,1;1,1,1,1)$ & $A^2$
\\[.3ex]
4&$(\bar2,\bar2,1,1;1,1,1,1)$ & $\varPhi_{\bar1\bar2}$
\\[.3ex]
1&$(1,1,2,\bar2;1,1,1,1)$& $\varPhi_{3\bar4}$
\\[.3ex]
1&$(1,1,\bar2,2,;1,1,1,1)$& $\varPhi_{\bar34}$
\\[.3ex]
12& $(1,1,1,1;1,1,1,1)$ & $A^3$
\\[.3ex]
12& $(1,1,1,1;1,1,1,1)$ & $A^4$
\\[.3ex]
4 & $(1,1,2,2;1,1,1,1)$ & $\varPhi_{34}$
\\[.3ex]
2&$(\bar2,1,1,1;4,1,1,1)$&
\\[.3ex]
2&$(1,\bar2,1,1;1,4,1,1)$&
\\[.3ex]
2&$(1,1,2,1;4,1,1,1)$&
\\[.3ex]
2&$(1,1,1,2;1,4,1,1)$&
\\[.3ex]
2&$(\bar2,1,1,1;1,1,4,1)$&
\\[.3ex]
2& $(1,\bar2,1,1;1,1,1,4)$&
\\[.3ex]
2&$(1,1,2,1;1,1,1,4)$&
\\[.3ex]
2&$ (1,1,1,2;1,1,4,1)$&
\\[.3ex]
1&$(1,1,1,1;6,1,1,1)$&
\\[.3ex]
1&$(1,1,1,1;1,6,1,1)$&
\\[.3ex]
2& $(1,1,1,1;4,4,1,1)$ &
\\[.3ex]
1&$(1,1,1,1;1,1,6,1)$&
\\[.3ex]
1&$(1,1,1,1;1,1,1,6)$&
\\[.3ex]
2& $(1,1,1,1;1,1,4,4)$ &
\\[.3ex]
1&$(1,1,1,1;4,1,4,1)$&
\\[.3ex]
1&$(1,1,1,1;4,1,1,4)$&
\\[.3ex]
1&$(1,1,1,1;1,4,4,1)$&
\\[.3ex]
1&$(1,1,1,1;1,4,1,4)$&
\\[.3ex]
\hline
\end{tabular}
\end{center}
\caption{Massless spectrum charged with respect to the $U(2)^2 \times U(2)^2 \times USp(4)^2 \times USp(4)^2$ gauge group, where the field $A^i$ is in the antisymmetric representation of the corresponding $SU(2)^i$ factor. To lighten the notation we have not explicitly listed the $U(1)$ charges. They can be easily derived from the $U(2)$ representations.}
\label{spectrum3}
\end{table}

The Abelian anomaly matrix is
\begin{equation}
C_{ab}= \frac{4}{\pi^2}\,
\left(
  \begin{array}{cccc}
    -3 & -1 & 0 & 0 \\
    -1 & -3 & 0 & 0 \\
    0 & 0 & 3 & 1 \\
    0 & 0 & 1 & 3 \\
  \end{array}
\right) \,,
\end{equation}
As usual, the gauge anomalies $C_{ab}$ are
cancelled by the variation of the gauge kinetic functions
\begin{equation}
\begin{split}
&f^{(1)}_1 = S+T_1+2T_2-2T_3+\alpha_1\,(2M_1+M_2)+\alpha_2\, M_3 \,,
\\
&f^{(1)}_2 = S+T_1+2T_2-2T_3+\alpha_1\, (2M_1-M_2) - \alpha_2\, M_3 \,,
\\
&f^{(2)}_1 = S+T_1+2T_2-2T_3 -\alpha_1 \, (2M_1+M_2)+\alpha_2 \, M_3 \,,
\\
&f^{(2)}_2 = S+T_1+2T_2-2T_3-\alpha_1 \, (2M_1-M_2) - \alpha_2 \, M_3 \,,
\\
\end{split}
\end{equation}
under the non-linear transformations of the closed-string fields
\begin{equation}
\begin{split}
&\delta
S=4\,\lambda_S\left[-\varLambda^{(1)}_1-\varLambda^{(1)}_2+\varLambda^{(2)}_1+\varLambda^{(2)}_2\right] \,,
\\
&\delta T_1=2\,\delta T_2=-2\,\delta
T_3=4\,\lambda\left[\varLambda^{(1)}_1+\varLambda^{(1)}_2-\varLambda^{(2)}_1-\varLambda^{(2)}_2\right] \,,
\\
&\delta M_1=2\lambda_1\,\alpha_1
\left[\varLambda^{(1)}_1+\varLambda^{(1)}_2+\varLambda^{(2)}_1+\varLambda^{(2)}_2\right] \,,
\\
&\delta M_2=2\,\lambda_2\,\alpha_1
\left[\varLambda^{(1)}_1-\varLambda^{(1)}_2+\varLambda^{(2)}_1-\varLambda^{(2)}_2\right] \,,
\\
&\delta M_3=2\,\lambda_3\,\alpha_2 \left[-\varLambda^{(1)}_1+\varLambda^{(1)}_2+\varLambda^{(2)}_1-\varLambda^{(2)}_2\right]\,.
\\
\end{split}
\end{equation}
Taking into account that the $U(1)$ generators are not canonically normalised, one gets the system of equations
\begin{equation}
\begin{split}
&C_{11}=C_{22}=-C_{33}=-C_{44}=4\left(-4\,\lambda_S+12\,\lambda+4\,\alpha_1^2\,\lambda_1+2\,\alpha_1^2\,
\lambda_2-2\,\alpha_2^2\,\lambda_3\right) \,,
\\
&C_{12}=C_{21}=-C_{34}=-C_{43}=4\left(-4\,\lambda_S+12\,\lambda+4\,\alpha_1^2\,\lambda_1-2\,\alpha_1^2\,
\lambda_2+2\,\alpha_2^2\,\lambda_3\right) \,,
\\
&C_{13}=C_{24}=-C_{31}=-C_{42}=4\left(4\,\lambda_S-12\,\lambda +4\,\alpha_1^2\,\lambda_1+2\,\alpha_1^2\,
\lambda_2+2\,\alpha_2^2\,\lambda_3 \right) \,,
\\
&C_{14}=C_{23}=-C_{32}=-C_{41}=4\left(4\,\lambda_S-12\,\lambda+4\,\alpha_1^2\,\lambda_1-2\,\alpha_1^2\,\lambda_2-2\,
\alpha_2^2\, \lambda_3\right) \,,
\\
\end{split}
\end{equation}
that also admit the solution (\ref{sys1}).

We can now turn to the analysis of non-perturbative effects induced by the ${\rm E}1_3$ branes, that are a distance away from the non-magnetised branes. For this reason, we shall restrict our attention only to rigid instantons associated to the fractional branes,
that we label by the vector $(k_1 , k_2 , k_3 , k_4)$.  As in the previous cases, we consider single instanton contributions, corresponding to branes with $SO(1)$ Chan-Paton group, so that $k_i =0,1$.  The zero-mode structure of these ${\rm E}1$ instantons is listed in table \ref{finalinst}.

\begin{table}[htdp]
\begin{center}
\begin{tabular}{|c|c|c|c|}
\hline
Instanton & $(k_1,k_2,k_3,k_4)$ & Representation & Zero mode
\\[1ex]
\hline
\hline
${\rm E}1_o$ & $(1,0,0,0)$ & $(1,2,1,1,1,1,1,1)$ & $\eta^{o}_i$
\\[1ex]
${\rm E}1_g$ & $(0,1,0,0)$ & $(2,1,1,1,1,1,1,1)$ & $\eta^{g}_i$
\\[1ex]
${\rm E}1_f$ & $(0,0,1,0)$ & $(1,1,\bar 2,1,1,1,1,1)$ & $\eta^{f}_i$
\\[1ex]
${\rm E}1_h$ & $(0,0,0,1)$ & $(1,1,1,\bar 2,1,1,1,1)$ & $\eta^{h}_i$
\\[1ex]
\hline
\end{tabular}
\end{center}
\caption{Charged zero-mode structure for $SO(1)$ E1 instantons for models with linear superpotential. The index $i$ refers to the (anti)fundamental representation of $U(2)$.}
\label{finalinst}
\end{table}

Let us analyse the case of a single ${\rm E}1_o$ instanton in some detail. The gauge-invariant instanton action including both neutral and charged zero modes is
\begin{equation}
S_{\rm inst} = S_{{\rm E}1_o} + \sum_{i,j=1}^2 \eta^{o}_i \, A^2_{ij}\, \eta^{o}_j +\sum_{i,j,k=1}^2\eta^{o}_i \, (\varPhi_{1\bar2})_{ki}\, (\varPhi_{\bar1\bar2})_{kj}\, \eta^{o}_j \,.
\end{equation}
Upon integration over the two charged zero modes $\eta^{o}_i$ one gets the following non-perturbative contribution to the superpotential
\begin{equation}
\mathcal{W}_{\rm non-pert} \, = \,  e^{- S_{{\rm E}1_o}} \, \sum_{i,j=1}^2\epsilon_{ij} \, \left[A_{ij}^2+\sum_{k=1}^2 (\varPhi_{1\bar2})_{ki}\, (\varPhi_{\bar1\bar2})_{kj}\right] \,.
\label{sys6}
\end{equation}

Similar results are obtained when considering the other instantons. For instance, an ${\rm E}1_g$ instanton induces the coupling
$A^1 + \varPhi_{\bar12}\, \varPhi_{\bar1\bar2}$, an ${\rm E}1_f$ instanton induces the coupling
$A^3+ \varPhi_{3\bar4}\, \varPhi_{34}$, and finally an ${\rm E}1_h$ instanton induces the coupling
$A^4+\varPhi_{\bar34}\, \Phi_{34}$. Equations   (\ref{variation4}), (\ref{action}),
(\ref{sys1}), (\ref{sys2}),  (\ref{sys3})  determine the gauge variations of the instanton actions
\begin{equation}
\delta S_{{\rm E}1_o} = 2\, \varLambda^{(1)}_2 \,, \quad
\delta S_{{\rm E}1_g} = 2\, \varLambda^{(1)}_1 \,, \quad
\delta S_{{\rm E}1_f} = -2\, \varLambda^{(2)}_1 \,, \quad
\delta S_{{\rm E}1_h}=-2\, \varLambda^{(2)}_2 \,,
\end{equation}
which imply, as expected, that the operators $e^{-S_{{\rm E}1}}\, (A+\varPhi\,\varPhi)$ are gauge invariant.

In the limit of rigid supersymmetry, and under the assumption that open-string moduli can be consistently frozen,
the linear terms in the superpotential induce Polony-like supersymmetry breaking. It is clear, however,
that open string moduli are dynamical and it is expected the non-magnetised D9 and D5 branes be unstable so that supersymmetry
is effectively restored. In a sense, the bulk branes are attracted towards the rigid instantons.

\section{Conclusions and Perspectives}

Stringy instantons play a crucial role in generating hierarchically small masses, low-energy supersymmetry breaking
and also conformal symmetry breaking.

In this paper, we have analysed in detail various instanton effects in the context of the simplest intersecting brane model with rigid cycles, {\it i.e.} the $\mathbb{Z}_2 \times \mathbb{Z}_2$ orientifold with discrete torsion. We have performed a complete
analysis based on anomaly cancelation, instantonic zero modes and the gauge invariance of the instanton-induced
superpotential terms. We have then built explicit examples where linear and/or quadratic terms in the superpotential are generated non-perturbatively, and may trigger supersymmetry breaking and moduli stabilisation. Moreover, we have also noticed how instantons can induce a low-scale breaking of conformal invariance, thus offering a possible solution to the hierarchy problem, as suggested in \cite{fv}.

However, we have only partially fulfilled our original goals.  One of our driving motivations was trying to connect non-tachyonic non-supersymmetric open-string models presented in \cite{aadds}, with supersymmetric vacua involving magnetised D9 branes on the same orbifold.
Our conjecture relied on the observations that in both models the closed-string sector has ${\cal N}=1$ supersymmetry, with identical spectra and configuration of orientifold planes. It is  plausible to believe that, even if classically stable, the non-supersymmetric solution \cite{aadds} is quantum-mechanically unstable, thus decaying into the supersymmetric solution \cite{cristina,bcms}.
This would be in agreement with what was recently argued to happen in an $\mathcal{N}=2$ set-up \cite{carlo2}.
However, although we lack a quantitative understanding of the dynamics, we cannot refrain from conjecturing the following scenario\footnote{E.D. is grateful to Valery Rubakov for enlightening discussions on this issue.}.
The non-supersymmetric orientifold model contains
stacks of 16 ${\rm D}5_{1,2}$ branes and 16 $\overline{{\rm D5}}_3$ antibranes, in addition to 16 D9 branes. All D5 branes
can be actually viewed  as zero-size gauge instantons on the D9 branes \cite{witten}. In fact, although in the supersymmetric case,
the instanton size is a modulus, in the case  of brane supersymmetry breaking, where supersymmetry is broken at the string scale,  the instanton size is not a modulus any longer. However, instantons would energetically prefer to expand to their maximum size within
the worldvolume of the D9 branes, and because of the conservation of topological charges, the original D5 (anti)branes are actually converted into diluted magnetic fluxes on the D9 branes. It is very tempting to believe that this dynamical process is triggered by some instanton in the orbifold geometry, even though the dynamics of the process is probably highly nontrivial.

Finally, although we concentrated on the case of single instantons with $SO(1)$ gauge group, so to have the minimal number (two) of
neutral zero modes  and thus be able to induce non-perturbative corrections to the superpotential, there is compelling evidence
that, even in the presence of a higher number of zero modes, non-perturbative (multi-instanton) effects can be generated
\cite{inaki, pablo,Billo':2008sp}. Also, there are possibly stable single E$3$ instantons in the model that might induce interesting effects.  It would be interesting to work out the S-dual version or the F-theory uplift of our configurations, so to exploit the dynamics of instantons beyond the naive minimal couplings with charged matter.

\vskip 24pt


\section*{Acknowledgments}

We  thank Pablo Camara and Valery Rubakov for enlightening discussions.  C.A. would like to thank the Laboratoire de Physique Theorique of Ecole Normale Superieure and the Centre de Physique Theorique of Ecole Polytechnique for their warm hospitality during the early stage of this collaboration.
The present research was supported in part by the European ERC Advanced Grant 226371 MassTeV, by the CNRS
PICS no. 3747 and 4172, in part by the grant ANR-05-BLAN-0079-02, in part by the RTN contracts MRTN-CT-2004-005104 and MRTN-CT-2004-503369, in part by the grant CNCSIS Ideas, and in part by the Italian MIUR-PRIN contract 20075ATT78.

\vskip 36pt


\appendix

\section{The $T^6/\mathbb{Z}_2\times \mathbb{Z}_2$ orbifolds and its basic amplitudes}

In this first Appendix we review the basic definitions of the $T^6/\mathbb{Z}_2 \times \mathbb{Z}_2$ orbifold and list the basic holomorphic amplitudes and their space-time interpretation.

We assume, for simplicity, that the six-torus factorises as $T^6 = T^2_1 \times T^2_2 \times T^2_3$, with complex coordinates
$z_i $ on each $T^2_i$. The $\mathbb{Z}_2 \times \mathbb{Z}_2$ is then generated by $g$, $f$ and $h$, whose action on the compact coordinates is
\begin{equation}
\begin{split}
&g\,:  \qquad z_1 \to +z_1\,, \quad z_2 \to - z_2 \,, \quad z_3 \to -z_3\,,
\\
&f\,: \qquad z_1 \to -z_1\,, \quad z_2 \to + z_2 \,, \quad z_3 \to -z_3\,,
\\
&h\,: \qquad z_1 \to -z_1\,, \quad z_2 \to - z_2 \,, \quad z_3 \to +z_3\,.
\\
\end{split}
\label{agenerators}
\end{equation}
We can conveniently define the $3\times 3$ diagonal matrices $\lambda_\ell = {\rm diag}\,(e^{2i\pi \lambda^1_\ell} \,,\, e^{2i\pi \lambda^2_\ell} \,,\, e^{2i\pi \lambda^3_\ell} )$, one for each of the $\mathbb{Z}_2 \times \mathbb{Z}_2$ generators,
$\ell = g ,\ f,\ h$, so that the action of the $\gamma_\ell$ generator on the complex coordinate $z = (z_1 \,,\, z_2 \,,\, z_3 )$ is given by
\begin{equation}
\gamma_\ell \,:\qquad z \to \lambda_\ell \cdot z = (e^{2i\pi \lambda_\ell^1} \, z_1\,,\, e^{2i\pi \lambda_\ell^2} \, z_2\,,\, e^{2i\pi \lambda_\ell^3} \, z_3),
\label{agamma}
\end{equation}
Comparison with eq. (\ref{agenerators}) properly defines the various $\lambda_\ell^i$. It will be useful to associate also a matrix $\lambda_o$ to the identity $o$ of the orbifold group, so that $\lambda_o^i =0$ for all $i$'s.

The action of the orbifold group on the world-sheet fermions is dictated by eq. (\ref{agenerators}) or (\ref{agamma}) together with the requirement that the two-dimensional supercurrent be invariant. The contribution of the world-sheet fermions to the various one-loop vacuum amplitudes is given, as usual, by a combination of Jacobi theta functions, whose characteristics depend on the orbifold (un)twisted sector and on the insertion of projection operators. More concretely, for a $\gamma_\mu$ twisted sector with the insertion in the trace of the $\gamma_\nu$ generator
\begin{equation}
\begin{split}
T_{\mu\nu} (\zeta) & \equiv
{\rm tr}_{\mathcal{H}_\mu} \left( \gamma_\nu\, q^{L_0 - c/24} \right)
\\
&= \frac{1}{2} \sum_{\alpha , \beta=0,\frac{1}{2}} C_{\alpha \beta}\,
\frac{\eta}{\theta \left[ {\alpha \atop \beta}\right]}
\frac{\theta^2 \left[ {\alpha \atop\beta}\right]
\, \theta \left[ {\alpha +\lambda^1_\mu \atop\beta +\lambda^1_\nu}\right] (\zeta_1|\tau)
\, \theta \left[ {\alpha +\lambda^2_\mu \atop\beta +\lambda^2_\nu}\right] (\zeta_2|\tau)
\, \theta \left[ {\alpha +\lambda^3_\mu \atop\beta +\lambda^3_\nu}\right] (\zeta_3|\tau)}{\eta^5} \,,
\\
\end{split}
\label{Tmunu}
\end{equation}
where we assume that both upper and lower characteristics of the theta functions are defined mod 1.
This equation needs some comments. The coefficients $C_{\alpha \beta}$ are the relative phases induced by the GSO projection, that we have implicitly included into the trace. For the type IIB superstring they are given by
\begin{equation}
C_{\alpha \beta} = e^{2 i \pi (\alpha + \beta + 2 \alpha\beta)}\,,
\end{equation}
for the holomorphic and anti-holomorphic sectors. The world-sheet bosonic ghosts, associated to super-reparametrisations, contribute with the term $\eta/\theta$, while the remaining theta functions correspond to the world-sheet fermions.
Typically, the ghost contributions cancel exactly those from the two light-cone fermions, however we have preferred to write them explicitly in eq. (\ref{Tmunu}) since this is not any more the case for Euclidean-brane instantons. Finally, we have allowed for a non-trivial $\zeta$ dependence of the theta functions originating from the internal fermions, since these will emerge when discussing open-string amplitudes in the presence of magnetic backgrounds.

To derive the spectrum of light excitations, it is convenient to write the $T_{\mu\nu}$ in terms of the $SO(2n)$ characters defined for instance in \cite{carlo, emilian, massimo}. Typically, the partition function counts only the physical degrees of freedom that, for massless states, are associated to the representations of the little group $SO(8)$. However, in what follows we find useful to use a covariant description, barring in mind that ghosts precisely cancel the (un-physical) longitudinal polarisations. As a result, we can write for instance
\begin{equation}
\begin{split}
T_{oo}
&=  V_4O_2O_2O_2+O_4O_2V_2O_2+O_4O_2O_2V_2+O_4V_2O_2O_2
\\
&+  O_4V_2V_2V_2 + V_4O_2V_2V_2+V_4V_2V_2O_2+V_4V_2O_2V_2
\\
&- ( S_4 S_2 S_2 S_2 + S_4 S_2C_2C_2 + S_4C_2S_2C_2 + S_4C_2C_2S_2
\\
&+ C_4S_2S_2C_2 + C_4S_2C_2S_2 + C_4C_2S_2S_2 + C_4C_2C_2C_2 )\, q^{-\frac{1}{8}}\,,
\\
\end{split}
\end{equation}

\begin{equation}
\begin{split}
T_{og}
&= V_4O_2O_2O_2 -O_4O_2V_2O_2-O_4O_2O_2V_2+O_4V_2O_2O_2
\\
&+O_4V_2V_2V_2 +V_4O_2V_2V_2-V_4V_2V_2O_2-V_4V_2O_2V_2
\\
&- (S_4S_2S_2S_2+S_4S_2C_2C_2-S_4C_2S_2C_2-S_4C_2C_2S_2
\\
&-C_4S_2S_2C_2-C_4S_2C_2S_2+C_4C_2S_2S_2+C_4C_2C_2C_2)\, q^{-\frac{1}{8}} \,,
\\
\end{split}
\end{equation}

\begin{equation}
\begin{split}
T_{of}
&= V_4O_2O_2O_2 + O_4O_2V_2O_2-O_4O_2O_2V_2-O_4V_2O_2O_2
\\
&+O_4V_2V_2V_2 -V_4O_2V_2V_2-V_4V_2V_2O_2+V_4V_2O_2V_2
\\
& -( S_4S_2S_2S_2 - S_4S_2C_2C_2 + S_4C_2S_2C_2-S_4C_2C_2S_2
\\
&-C_4S_2S_2C_2+C_4S_2C_2S_2-C_4C_2S_2S_2+C_4C_2C_2C_2 )\, q^{-\frac{1}{8}}\,,
\\
\end{split}
\end{equation}

\begin{equation}
\begin{split}
T_{oh}
&= V_4O_2O_2O_2 -O_4O_2V_2O_2+O_4O_2O_2V_2-O_4V_2O_2O_2
\\
&+O_4V_2V_2V_2 -V_4O_2V_2V_2+V_4V_2V_2O_2-V_4V_2O_2V_2
\\
&- (S_4S_2S_2S_2-S_4S_2C_2C_2-S_4C_2S_2C_2+S_4C_2C_2S_2
\\
&+C_4S_2S_2C_2-C_4S_2C_2S_2-C_4C_2S_2S_2+C_4C_2C_2C_2) \, q^{-\frac{1}{8}}\,,
\end{split}
\label{consid}
\end{equation}
and similar expressions for the twisted amplitudes. The extra factor of $q$, whenever present, takes into account the contribution of the ghosts to the zero-point energy. As a result, not only terms like $V_4 O_2 O_2 O_2$ correspond to massless states, but also space-time fermions like $S_4S_2S_2S_2$ have vanishing mass, since the factor $q^{-\frac{1}{8}}$ compensates for the higher conformal weight of $SO(4)$ spinorial characters.

\subsection{Amplitudes for the Euclidean branes}

When dealing with Euclidean brane instantons in a given D-brane and O-plane vacuum configuration, one has to be careful with open-string frequencies, since on E branes the space-time coordinates obey Dirichlet boundary conditions, while on D branes they obey Neumann  boundary conditions. In this paper, we consider the effect of E5 instantons and three types of E1 instantons on a configuration with O9 and O5 planes, and (possibly magnetised) D9 and D5 branes. If we indicate by $x$ the non-compact space-time coordinates and by $z_i$ the three complex coordinates on the three $T^2$'s, we have the following sets of boundary conditions
\begin{equation}
\begin{array}{lcccc}
 &\ x\ &\ z_1\ &\ z_2\ &\ z_3
 \\
{\rm O}9 \quad & \times & \times & \times & \times
\\
{\rm O}5_1  \quad & \times & \times & \cdot & \cdot
\\
{\rm O}5_2 \quad & \times & \cdot & \times & \cdot
\\
{\rm O}5_3 \quad & \times & \cdot & \cdot & \times
\\
{\rm D}9 \quad & \times & \times & \times & \times
\\
{\rm E}5\quad & \cdot & \times & \times & \times
\\
{\rm E}1_1\quad & \cdot & \times & \cdot & \cdot
\\
{\rm E1}_2 \quad & \cdot & \cdot & \times & \cdot
\\
{\rm E}1_3 \quad & \cdot & \cdot & \cdot & \times
\\
{\rm D}5_1 \quad & \times & \times & \cdot & \cdot
\end{array}
\end{equation}
where a cross (dot) indicates that the object wraps (is localised along) the corresponding directions. This table suggests to assign a vector $\rho^M = (\rho^M_0 \,,\,\rho^M_1\,,\,\rho^M_2\, ,\, \rho^M_3)$ to each object in the list, with $M=1,\ldots , 10$ corresponding to the ordered sequence O9, $\ldots$, ${\rm D}5_1$. Then $\rho^M_\alpha =0$ ($\rho^M_\alpha =1$) if the $M$-th object wraps (is localised along) the $\alpha$-th directions, while
\begin{equation}
2 \, \omega^{MN} = \rho^M + \rho^N \ {\rm mod} \ 2
\end{equation}
clearly counts the relative ``Neumann-Dirichlet boundary conditions''. As a result, the associated open-string world-sheet fermions and superghosts contribute to the partition function with
\begin{equation}
T^{MN}_{o\nu} = \frac{1}{2} \sum_{\alpha , \beta = 0 , \frac{1}{2}} C_{\alpha\beta} \frac{\eta}{\theta \left[ {\alpha \atop \beta}\right]}
\frac{
\theta^2 \left[ {\alpha + \omega^{MN}_0 \atop \beta}\right]\,
\theta \left[ {\alpha + \omega^{MN}_1 \atop \beta+\lambda_\nu^1}\right]\,
\theta \left[ {\alpha + \omega^{MN}_2 \atop \beta+\lambda_\nu^2}\right]\,
\theta \left[ {\alpha + \omega^{MN}_3 \atop \beta+\lambda_\nu^3}\right]\,
}{\eta^5}\,,
\end{equation}
where, as before, we assume that upper and lower characteristics in the theta functions are defined mod 1. Clearly, whenever $M=N$, or $\omega^{MN} \equiv 0$ the $T^{MN}_{o\nu}$ reduce to the $T_{o\nu}$ defined in eq. (\ref{Tmunu}). In the following subsections, we give the character expansions of some relevant $T^{MN}_{o\nu}$.

\subsection{$T^{MN}_{o\nu}$ for open strings between E5 and D9 branes}

To derive the spectrum of light excitations, it is convenient to write the $T^{56}_{o\nu}$ in terms of the $SO(2n)$ characters defined for instance in \cite{carlo}. One has
\begin{equation}
\begin{split}
T^{56}_{oo}
&= S_4O_2O_2O_2+S_4O_2V_2V_2+S_4V_2O_2V_2+S_4V_2V_2O_2
\\
&+C_4O_2O_2V_2+C_4O_2V_2O_2+C_4V_2O_2O_2+C_4V_2V_2V_2
\\
&- (O_4S_2S_2C_2 + O_4S_2C_2S_2 + O_4C_2S_2S_2 + O_4C_2C_2C_2
\\
& + V_4S_2S_2S_2 + V_4S_2C_2C_2 + V_4C_2S_2C_2 + V_4C_2C_2S_2) \, q^{-\frac{1}{8}}\,,
\\
\end{split}
\end{equation}

\begin{equation}
\begin{split}
T^{56}_{og}
&=S_4O_2O_2O_2+S_4O_2V_2V_2-S_4V_2O_2V_2-S_4V_2V_2O_2
\\
&-C_4O_2O_2V_2-C_4O_2V_2O_2+C_4V_2O_2O_2+C_4V_2V_2V_2
\\
&- ( - O_4S_2S_2C_2 - O_4S_2C_2S_2 + O_4C_2S_2S_2 + O_4C_2C_2C_2
\\
&+ V_4S_2S_2S_2 + V_4S_2C_2C_2 - V_4C_2S_2C_2 - V_4C_2C_2S_2) \, q^{-\frac{1}{8}}\,,
\\
\end{split}
\end{equation}

\begin{equation}
\begin{split}
T^{56}_{of}
&= S_4O_2O_2O_2-S_4O_2V_2V_2+S_4V_2O_2V_2-S_4V_2V_2O_2
\\
&-C_4O_2O_2V_2+C_4O_2V_2O_2-C_4V_2O_2O_2+C_4V_2V_2V_2
\\
&-(-O_4S_2S_2C_2+O_4S_2C_2S_2-O_4C_2S_2S_2+O_4C_2C_2C_2
\\
&+V_4S_2S_2S_2-V_4S_2C_2C_2+V_4C_2S_2C_2-V_4C_2C_2S_2 ) \, q^{-\frac{1}{8}}\,,
\\
\end{split}
\end{equation}

\begin{equation}
\begin{split}
T^{56}_{oh}
&=S_4O_2O_2O_2-S_4O_2V_2V_2-S_4V_2O_2V_2+S_4V_2V_2O_2
\\
&+C_4O_2O_2V_2-C_4O_2V_2O_2-C_4V_2O_2O_2+C_4V_2V_2V_2
\\
&-( O_4S_2S_2C_2-O_4S_2C_2S_2-O_4C_2S_2S_2+O_4C_2C_2C_2
\\
&+V_4S_2S_2S_2-V_4S_2C_2C_2-V_4C_2S_2C_2+V_4C_2C_2S_2 ) \, q^{-\frac{1}{8}}\,.
\\
\end{split}
\end{equation}
Considerations similar to those given after eq. (\ref{consid}) apply also here for a proper interpretation of the previous expressions.

\subsection{$T^{MN}_{o\nu}$ for open strings between ${\rm E}1_1$ and D9 branes}

To derive the spectrum of light excitations, it is convenient to write the $T^{57}_{o\nu}$ in terms of the $SO(2n)$ characters defined for instance in \cite{carlo}. One has
\begin{equation}
\begin{split}
T^{57}_{oo}
&= S_4O_2S_2C_2+S_4O_2C_2S_2+S_4V_2S_2S_2+S_4V_2C_2C_2
\\
&+C_4O_2S_2S_2+C_4O_2C_2C_2+C_4V_2S_2C_2+C_4V_2C_2S_2
\\
&- (O_4S_2O_2O_2+O_4S_2V_2V_2+O_4C_2O_2V_2+O_4C_2V_2O_2
\\
&+V_4S_2O_2V_2+V_4S_2V_2O_2+V_4C_2O_2O_2+V_4C_2V_2V_2) \, q^{-\frac{1}{8}} \,,
\\
\end{split}
\end{equation}

\begin{equation}
\begin{split}
T^{57}_{og}
&=-S_4O_2S_2C_2-S_4O_2C_2S_2+S_4V_2S_2S_2+S_4V_2C_2C_2
\\
&+C_4O_2S_2S_2+C_4O_2C_2C_2-C_4V_2S_2C_2-C_4V_2C_2S_2
\\
&- (O_4S_2O_2O_2+O_4S_2V_2V_2-O_4C_2O_2V_2-O_4C_2V_2O_2
\\
&-V_4S_2O_2V_2-V_4S_2V_2O_2+V_4C_2O_2O_2+V_4C_2V_2V_2) \, q^{-\frac{1}{8}} \,,
\\
\end{split}
\end{equation}

\begin{equation}
\begin{split}
T^{57}_{of}
&=i (S_4O_2S_2C_2-S_4O_2C_2S_2+S_4V_2S_2S_2-S_4V_2C_2C_2
\\
&-C_4O_2S_2S_2+C_4O_2C_2C_2-C_4V_2S_2C_2+C_4V_2C_2S_2)
\\
&-i (O_4S_2O_2O_2-O_4S_2V_2V_2+O_4C_2O_2V_2-O_4C_2V_2O_2
\\
&-V_4S_2O_2V_2+V_4S_2V_2O_2-V_4C_2O_2O_2+V_4C_2V_2V_2) \, q^{-\frac{1}{8}} \,,
\\
\end{split}
\end{equation}

\begin{equation}
\begin{split}
T^{57}_{oh}
&=i (-S_4O_2S_2C_2+S_4O_2C_2S_2+S_4V_2S_2S_2-S_4V_2C_2C_2
\\
&-C_4O_2S_2S_2+C_4O_2C_2C_2+C_4V_2S_2C_2-C_4V_2C_2S_2)
\\
&-i (O_4S_2O_2O_2-O_4S_2V_2V_2-O_4C_2O_2V_2+O_4C_2V_2O_2
\\
&+V_4S_2O_2V_2-V_4S_2V_2O_2-V_4C_2O_2O_2+V_4C_2V_2V_2) \, q^{-\frac{1}{8}} \,.
\\
\end{split}
\end{equation}
Considerations similar to those given after eq. (\ref{consid}) apply also here for a proper interpretation of the previous expressions.

\subsection{$T^{MN}_{o\nu}$ for open strings between ${\rm E}1_2$ and D9 branes}

To derive the spectrum of light excitations, it is convenient to write the $T^{58}_{o\nu}$ in terms of the $SO(2n)$ characters defined for instance in \cite{carlo}. One has
\begin{equation}
\begin{split}
T^{58}_{oo}
&= S_4S_2O_2C_2+S_4C_2O_2S_2+S_4S_2V_2S_2+S_4C_2V_2C_2
\\
&+C_4S_2O_2S_2+C_4C_2O_2C_2+C_4S_2V_2C_2+C_4C_2V_2S_2
\\
&- (O_4O_2S_2O_2+O_4V_2S_2V_2+O_4O_2C_2V_2+O_4V_2C_2O_2
\\
&+V_4O_2S_2V_2+V_4V_2S_2O_2+V_4O_2C_2O_2+V_4V_2C_2V_2) \, q^{-\frac{1}{8}} \,,
\\
\end{split}
\end{equation}

\begin{equation}
\begin{split}
T^{58}_{og}
&=i (S_4S_2O_2C_2-S_4C_2O_2S_2+S_4S_2V_2S_2-S_4C_2V_2C_2
\\
&-C_4S_2O_2S_2+C_4C_2O_2C_2-C_4S_2V_2C_2+C_4C_2V_2S_2)
\\
&-i (O_4O_2S_2O_2-O_4V_2S_2V_2+O_4O_2C_2V_2-O_4V_2C_2O_2
\\
&-V_4O_2S_2V_2+V_4V_2S_2O_2-V_4O_2C_2O_2+V_4V_2C_2V_2) \, q^{-\frac{1}{8}} \,,
\\
\end{split}
\end{equation}

\begin{equation}
\begin{split}
T^{58}_{of}
&= -S_4S_2O_2C_2-S_4C_2O_2S_2+S_4S_2V_2S_2+S_4C_2V_2C_2
\\
&+C_4S_2O_2S_2+C_4C_2O_2C_2-C_4S_2V_2C_2-C_4C_2V_2S_2
\\
&- (O_4O_2S_2O_2+O_4V_2S_2V_2-O_4O_2C_2V_2-O_4V_2C_2O_2
\\
&-V_4O_2S_2V_2-V_4V_2S_2O_2+V_4O_2C_2O_2+V_4V_2C_2V_2) \, q^{-\frac{1}{8}} \,,
\\
\end{split}
\end{equation}

\begin{equation}
\begin{split}
T^{58}_{oh}
&=i (S_4S_2O_2C_2-S_4C_2O_2S_2-S_4S_2V_2S_2+S_4C_2V_2C_2
\\
&+C_4S_2O_2S_2-C_4C_2O_2C_2-C_4S_2V_2C_2+C_4C_2V_2S_2)
\\
&-i (-O_4O_2S_2O_2+O_4V_2S_2V_2+O_4O_2C_2V_2-O_4V_2C_2O_2
\\
&-V_4O_2S_2V_2+V_4V_2S_2O_2+V_4O_2C_2O_2-V_4V_2C_2V_2) \, q^{-\frac{1}{8}} \,.
\\
\end{split}
\end{equation}
Considerations similar to those given after eq. (\ref{consid}) apply also here for a proper interpretation of the previous expressions.

\subsection{$T^{MN}_{o\nu}$ for open strings between ${\rm E}1_3$ and D9 branes}

To derive the spectrum of light excitations, it is convenient to write the $T^{59}_{o\nu}$ in terms of the $SO(2n)$ characters defined for instance in \cite{carlo}. One has
\begin{equation}
\begin{split}
T_{oo}^{59}
&= S_4S_2C_2O_2+S_4C_2S_2O_2+S_4S_2S_2V_2+S_4C_2C_2V_2
\\
&+C_4S_2S_2O_2+C_4C_2C_2O_2+C_4S_2C_2V_2+C_4C_2S_2V_2
\\
&-(O_4O_2O_2S_2+O_4V_2V_2S_2+O_4O_2V_2C_2+O_4V_2O_2C_2
\\
&+V_4O_2V_2S_2+V_4V_2O_2S_2+V_4O_2O_2C_2+V_4V_2V_2C_2) \, q^{-\frac{1}{8}} \,,
\\
\end{split}
\end{equation}

\begin{equation}
\begin{split}
T_{og}^{59}
&=i (-S_4S_2C_2O_2+S_4C_2S_2O_2-S_4S_2S_2V_2+S_4C_2C_2V_2
\\
&+C_4S_2S_2O_2-C_4C_2C_2O_2+C_4S_2C_2V_2-C_4C_2S_2V_2)
\\
&-i (-O_4O_2O_2S_2+O_4V_2V_2S_2-O_4O_2V_2C_2+O_4V_2O_2C_2
\\
&+V_4O_2V_2S_2-V_4V_2O_2S_2+V_4O_2O_2C_2-V_4V_2V_2C_2)\, q^{-\frac{1}{8}} \,,
\\
\end{split}
\end{equation}

\begin{equation}
\begin{split}
T_{of}^{59}
&=i (S_4S_2C_2O_2-S_4C_2S_2O_2-S_4S_2S_2V_2+S_4C_2C_2V_2
\\
&+C_4S_2S_2O_2-C_4C_2C_2O_2-C_4S_2C_2V_2+C_4C_2S_2V_2)
\\
&-i (-O_4O_2O_2S_2+O_4V_2V_2S_2+O_4O_2V_2C_2-O_4V_2O_2C_2
\\
&-V_4O_2V_2S_2+V_4V_2O_2S_2+V_4O_2O_2C_2-V_4V_2V_2C_2) \, q^{-\frac{1}{8}} \,,
\\
\end{split}
\end{equation}

\begin{equation}
\begin{split}
T_{oh}^{59}
&= -S_4S_2C_2O_2-S_4C_2S_2O_2+S_4S_2S_2V_2+S_4C_2C_2V_2
\\
&+C_4S_2S_2O_2+C_4C_2C_2O_2-C_4S_2C_2V_2-C_4C_2S_2V_2
\\
&-(O_4O_2O_2S_2+O_4V_2V_2S_2-O_4O_2V_2C_2-O_4V_2O_2C_2
\\
&-V_4O_2V_2S_2-V_4V_2O_2S_2+V_4O_2O_2C_2+V_4V_2V_2C_2) \, q^{-\frac{1}{8}} \,.
\\
\end{split}
\end{equation}
Considerations similar to those given after eq. (\ref{consid}) apply also here for a proper interpretation of the previous expressions.

\subsection{$\hat T^*_{o\nu}$ for open strings between E branes and O planes}

The contribution of instantonic branes to the M\"obius strip amplitude can be expressed in terms of
\begin{equation}
\hat T ^*_{o\nu} = \frac{1}{2} \sum_{\alpha ,\beta =0,\frac{1}{2}} C_{\alpha\beta} \frac{\hat\eta}{\hat\theta \left[ {\alpha \atop \beta}\right]} \frac{\hat\theta \left[ {\alpha \atop \beta+\frac{1}{2}}\right]^2 }{\hat\eta^2}
\frac{
\hat\theta \left[ {\alpha \atop \beta+\lambda_\nu^1}\right]
\hat\theta \left[ {\alpha \atop \beta+\lambda_\nu^2}\right]
\hat\theta \left[ {\alpha \atop \beta+\lambda_\nu^3}\right]}{\hat\eta^3}\,,
\end{equation}
where, as usual, the characteristics of the theta functions are defined mod 1.

To derive the spectrum of light excitations, it is convenient to write the $\hat T^*_{o\nu}$ in terms of the $SO(2n)$ characters defined for instance in \cite{carlo}. One has\begin{equation}
\begin{split}
\hat{T}_{oo}^*
&= O_4O_2O_2V_2+O_4O_2V_2O_2+O_4V_2O_2O_2+O_4V_2V_2V_2
\\
&-V_4O_2O_2O_2-V_4O_2V_2V_2-V_4V_2O_2V_2-V_4V_2V_2O_2
\\
&-(S_4S_2S_2S_2+S_4S_2C_2C_2+S_4C_2S_2C_2+S_4C_2C_2S_2
\\
&-C_4S_2S_2C_2-C_4S_2C_2S_2-C_4C_2S_2S_2-C_4C_2C_2C_2) \, q^{-\frac{1}{8}} \,,
\\
\end{split}
\end{equation}
\begin{equation}
\begin{split}
\hat{T}_{og}^*
&=-O_4O_2O_2V_2-O_4O_2V_2O_2+O_4V_2O_2O_2+O_4V_2V_2V_2
\\
&-V_4O_2O_2O_2-V_4O_2V_2V_2+V_4V_2O_2V_2+V_4V_2V_2O_2
\\
&-(S_4S_2S_2S_2+S_4S_2C_2C_2-S_4C_2S_2C_2-S_4C_2C_2S_2
\\
&+C_4S_2S_2C_2+C_4S_2C_2S_2-C_4C_2S_2S_2-C_4C_2C_2C_2) \, q^{-\frac{1}{8}} \,,
\\
\end{split}
\end{equation}
\begin{equation}
\begin{split}
\hat{T}_{of}^*
&= -O_4O_2O_2V_2+O_4O_2V_2O_2-O_4V_2O_2O_2+O_4V_2V_2V_2
\\
&-V_4O_2O_2O_2+V_4O_2V_2V_2-V_4V_2O_2V_2+V_4V_2V_2O_2
\\
&-(S_4S_2S_2S_2-S_4S_2C_2C_2+S_4C_2S_2C_2-S_4C_2C_2S_2
\\
&+C_4S_2S_2C_2-C_4S_2C_2S_2+C_4C_2S_2S_2-C_4C_2C_2C_2)\, q^{-\frac{1}{8}} \,,
\\
\end{split}
\end{equation}
\begin{equation}
\begin{split}
\hat{T}_{oh}^*
&= O_4O_2O_2V_2-O_4O_2V_2O_2-O_4V_2O_2O_2+O_4V_2V_2V_2
\\
&-V_4O_2O_2O_2+V_4O_2V_2V_2+V_4V_2O_2V_2-V_4V_2V_2O_2
\\
&-(S_4S_2S_2S_2-S_4S_2C_2C_2-S_4C_2S_2C_2+S_4C_2C_2S_2
\\
&-C_4S_2S_2C_2+C_4S_2C_2S_2+C_4C_2S_2S_2-C_4C_2C_2C_2) \, q^{-\frac{1}{8}} \,.
\\
\end{split}
\end{equation}
Considerations similar to those given after eq. (\ref{consid}) apply also here for a proper interpretation of the previous expressions.

\section{Partition function for open string in the presence of background magnetic fields}

We reproduce here the general partition function of magnetised D9 branes on a $T^6 /\mathbb{Z}_2 \times \mathbb{Z}_2$ orbifold with discrete torsion. $p_a$ and $q_\alpha$ denote different stacks of branes, while the intersection numbers and $S^{AB}_i$ are defined in eqs. (\ref{intersections}), (\ref{ointersections}) and (\ref{fixedintersections}). The magnetic field deformation on each $T^2$
is encoded in $z^A_i$, while $z^{AB}_i = z_i^A - z_i^B$ and $z^{AB'}_i = z_i^A + z_i^B$. For details about the construction of the annulus and M\"obius strip amplitudes see refs. \cite{magnetic, intersecting, cristina, carlo, SSmagnetic}.

The spectrum of open strings stretched between a ${\rm D}9^A$ brane and itself or its image ${\rm D}9^{A'}$ is encoded in the
annulus amplitude
\begin{equation}
\begin{split}
\mathcal{A}^{A,A^{(\prime)}}
&=\frac{1}{4}\int_0^{\infty}\frac{dt}{t^3}\Bigg\lbrace  p_a\bar p_a \Bigg[\tilde
P_1\tilde P_2\tilde P_3T_{oo}(0)
+\left(\tilde
P_1T_{og}(0)
+\tilde P_2T_{of}(0)+\tilde P_3T_{oh}(0)\right)\left(\frac{2\eta}{\theta_2}\right)^2\Bigg]
\\
&+I^{aa'}\left[\frac{p_a^2}{2}T_{oo}(2z_i^a\tau)+\frac{\bar
p_a^2}{2}T_{oo}(-2z_i^a\tau)\right]\prod_{i=1}^3\frac{i\eta}{\theta_1(2z_i^a\tau)}
\\
&-4I_1^{aa'}\left[\frac{p_a^2}{2}T_{og}(2z_i^a\tau)+\frac{\bar
p_a^2}{2}T_{og}(-2z_i^a\tau)\right]\frac{i\eta}{\theta_1(2z_1^a\tau)}\prod_{i=2,3}\frac{\eta}{\theta_2(2z_i^a\tau)}
\\
&-4I_2^{aa'}\left[\frac{p_a^2}{2}T_{of}(2z_i^a\tau)+\frac{\bar
p_a^2}{2}T_{of}(-2z_i^a\tau)\right]\frac{i\eta}{\theta_1(2z_2^a\tau)}\prod_{i=1,3}\frac{\eta}{\theta_2(2z_i^a\tau)}
\\
&+4I_3^{aa'}\left[\frac{p_a^2}{2}T_{oh}(2z_i^a\tau)+\frac{\bar
p_a^2}{2}T_{oh}(-2z_i^a\tau)\right]\frac{i\eta}{\theta_1(2z_3^a\tau)}\prod_{i=1,2}\frac{\eta}{\theta_2(2z_i^a\tau)}
\\
& +(a,a'\rightarrow\alpha,\alpha ')
\Bigg\rbrace\frac{1}{\eta^2}\\
\end{split}
\end{equation}
and in the M\"obius-strip amplitude
\begin{equation}
\begin{split}
\mathcal{M}
&=-\frac{1}{4}\int_0^{\infty}\frac{dt}{t^3}\Bigg\lbrace
\prod_{i=1}^3(m_i^{(a)})\left[p_a\hat T_{oo}(2z_i^a\tau)+\bar
p_a\hat T_{oo}(-2z_i^a\tau)\right]
\prod_{i=1}^3\frac{i\hat{\eta}}{\hat{\theta}_1(2z_i^a\tau)}
\\
&-\epsilon_1\left[p_a\hat T_{og}(2z_i^a\tau)+\bar p_a\hat
T_{og}(-2z_i^a\tau)\right]
\frac{im_1^{(a)}\hat{\eta}}{\hat{\theta}_1(2z_1^a\tau)}\prod_{i=2,3}\frac{n_i^{(a)}\hat{\eta}}{\hat{\theta}_2(2z_i^a\tau)}
\\
&-\epsilon_2\left[p_a\hat T_{of}(2z_i^a\tau)+\bar p_a\hat
T_{of}(-2z_i^a\tau)\right]
\frac{im_2^{(a)}\hat{\eta}}{\hat{\theta}_1(2z_2^a\tau)}\prod_{i=1,3}\frac{n_i^{(a)}\hat{\eta}}{\hat{\theta}_2(2z_i^a\tau)}
\\
&-\epsilon_3\left[p_a\hat T_{oh}(2z_i^a\tau)+\bar p_a\hat
T_{oh}(-2z_i^a\tau)\right]
\frac{im_3^{(a)}\hat{\eta}}{\hat{\theta}_1(2z_3^a\tau)}\prod_{i=1,2}\frac{n_i^{(a)}\hat{\eta}}{\hat{\theta}_2(2z_i^a\tau)}
\\
&+(a\rightarrow\alpha)
\Bigg\rbrace\frac{1}{\eta^2} \,.
\end{split}
\end{equation}
Here $\tilde P_i$ denote ``boosted'' compactification lattices
obtained replacing Kaluza-Klein momenta $k_i$ in the i-{\it th} torus
by $k_i\rightarrow k_i/\sqrt{n_i^2+(m_i^2/v_i^2)}$.

Oriented open strings, stretched between different stacks of branes, yield the following contributions to the annulus amplitude
\begin{equation}
\begin{split}
\mathcal{A}^{a,b^{(\prime)}}&=\frac{1}{4}\int_0^{\infty}\frac{dt}{t^3}\frac{1}{\eta^2} \Bigg\lbrace
I^{ab}\left[p_a\bar p_bT_{oo}(z_i^{ab}\tau)+\bar p_ap_bT_{oo}(-z_i^{ab}\tau)\right]\prod_{i=1}^3\frac{i\eta}{\theta_1(z_i^{ab}\tau)}
\\
&+I^{ab'}\left[p_ap_bT_{oo}(z_i^{ab'}\tau)+\bar p_a\bar p_bT_{oo}(-z_i^{ab'}\tau)\right]\prod_{i=1}^3\frac{i\eta}{\theta_1(z_i^{ab'}\tau)}
\\
&+S_g^{ab}I_1^{ab}\left[p_a\bar p_bT_{og}(z_i^{ab}\tau)+\bar
p_ap_bT_{og}(-z_i^{ab}\tau)\right]
\frac{i\eta}{\theta_1(z_1^{ab}\tau)}\prod_{i=2,3}\frac{\eta}{\theta_2(z_i^{ab}\tau)}
\\
&-S_g^{ab}I_1^{ab'}\left[p_ap_bT_{og}(z_i^{ab'}\tau)+\bar p_a\bar
p_bT_{og}(-z_i^{ab'}\tau)\right]
\frac{i\eta}{\theta_1(z_1^{ab'}\tau)}\prod_{i=2,3}\frac{\eta}{\theta_2(z_i^{ab'}\tau)}
\\
&+S_f^{ab}I_2^{ab}\left[p_a\bar p_bT_{of}(z_i^{ab}\tau)+\bar
p_ap_bT_{of}(-z_i^{ab}\tau)\right]
\frac{i\eta}{\theta_1(z_2^{ab}\tau)}\prod_{i=1,3}\frac{\eta}{\theta_2(z_i^{ab}\tau)}
\\
&-S_f^{ab}I_2^{ab'}\left[p_ap_bT_{of}(z_i^{ab'}\tau)+\bar p_a\bar
p_bT_{of}(-z_i^{ab'}\tau)\right]
\frac{i\eta}{\theta_1(z_2^{ab'}\tau)}\prod_{i=1,3}\frac{\eta}{\theta_2(z_i^{ab'}\tau)}
\\
&+S_h^{ab}I_3^{ab}\left[p_a\bar p_bT_{oh}(z_i^{ab}\tau)+\bar
p_ap_bT_{oh}(-z_i^{ab}\tau)\right]
\frac{i\eta}{\theta_1(z_3^{ab}\tau)}\prod_{i=1,2}\frac{\eta}{\theta_2(z_i^{ab}\tau)}
\\
&+S_h^{ab}I_3^{ab'}\left[p_ap_bT_{oh}(z_i^{ab'}\tau)+\bar p_a\bar
p_bT_{oh}(-z_i^{ab'}\tau)\right]
\frac{i\eta}{\theta_1(z_3^{ab'}\tau)}\prod_{i=1,2}\frac{\eta}{\theta_2(z_i^{ab'}\tau)}\Bigg\rbrace\,,
\\
\end{split}
\end{equation}

\begin{equation}
\mathcal{A}^{\alpha,\beta^{(\prime)}}=\mathcal{A}^{a,b^{(\prime)}}\quad {\rm with}\quad a,b, b'\rightarrow\alpha,\beta,\beta ' \,,
\end{equation}
and
\begin{equation}
\begin{split}
\mathcal{A}^{a,\alpha^{(\prime)}}&=\frac{1}{4}\int_0^{\infty}\frac{dt}{t^3} \frac{1}{\eta^2} \Bigg\lbrace
I^{a\alpha}\left[p_a\bar q_{\alpha}T_{oo}(z_i^{a\alpha}\tau)+\bar p_aq_{\alpha}T_{oo}(-z_i^{a\alpha}\tau)\right]\prod_{i=1}^3\frac{i\eta}{\theta_1(z_i^{a\alpha}\tau)}\\
&+I^{a\alpha'}\left[p_aq_{\alpha}T_{oo}(z_i^{a\alpha'}\tau)+\bar p_a\bar q_{\alpha}T_{oo}(-z_i^{a\alpha'}\tau)\right]\prod_{i=1}^3\frac{i\eta}{\theta_1(z_i^{a\alpha'}\tau)}\\
&+S_g^{a\alpha}I_1^{a\alpha}\left[p_a\bar
q_{\alpha}T_{og}(z_i^{a\alpha}\tau)+\bar
p_aq_{\alpha}T_{og}(-z_i^{a\alpha}\tau)\right]
\frac{i\eta}{\theta_1(z_1^{a\alpha}\tau)}\prod_{i=2,3}\frac{\eta}{\theta_2(z_i^{a\alpha}\tau)}\\
&-S_g^{a\alpha}I_1^{a\alpha'}\left[p_aq_{\alpha}T_{og}(z_i^{a\alpha'}\tau)+\bar
p_a\bar q_{\alpha}T_{og}(-z_i^{a\alpha'}\tau)\right]
\frac{i\eta}{\theta_1(z_1^{a\alpha'}\tau)}\prod_{i=2,3}\frac{\eta}{\theta_2(z_i^{a\alpha'}\tau)}\\
&-S_f^{a\alpha}I_2^{a\alpha}\left[p_a\bar
q_{\alpha}T_{of}(z_i^{a\alpha}\tau)+\bar
p_aq_{\alpha}T_{of}(-z_i^{a\alpha}\tau)\right]
\frac{i\eta}{\theta_1(z_2^{a\alpha}\tau)}\prod_{i=1,3}\frac{\eta}{\theta_2(z_i^{a\alpha}\tau)}\\
&+S_f^{a\alpha}I_2^{a\alpha'}\left[p_aq_{\alpha}T_{of}(z_i^{a\alpha'}\tau)+\bar
p_a\bar q_{\alpha}T_{of}(-z_i^{a\alpha'}\tau)\right]
\frac{i\eta}{\theta_1(z_2^{a\alpha'}\tau)}\prod_{i=1,3}\frac{\eta}{\theta_2(z_i^{a\alpha'}\tau)}\\
&-S_h^{a\alpha}I_3^{a\alpha}\left[p_a\bar
q_{\alpha}T_{oh}(z_i^{a\alpha}\tau)+\bar
p_aq_{\alpha}T_{oh}(-z_i^{a\alpha}\tau)\right]
\frac{i\eta}{\theta_1(z_3^{a\alpha}\tau)}\prod_{i=1,2}\frac{\eta}{\theta_2(z_i^{a\alpha}\tau)}\\
&-S_h^{a\alpha}I_3^{a\alpha'}\left[p_aq_{\alpha}T_{oh}(z_i^{a\alpha'}\tau)+\bar
p_a\bar q_{\alpha}T_{oh}(-z_i^{a\alpha'}\tau)\right]
\frac{i\eta}{\theta_1(z_3^{a\alpha'}\tau)}\prod_{i=1,2}\frac{\eta}{\theta_2(z_i^{a\alpha'}\tau)}
\Bigg\rbrace .\\
\end{split}
\end{equation}

\section{E-brane instanton partition functions and zero modes}

We shall focus here on the partition function of E-brane instantons in a vacuum configuration with magnetised D9 branes on a $T^6/\mathbb{Z}_2\times \mathbb{Z}_2 $ orbifold with discrete torsion. For simplicity we shall assume that identical magnetic fields are turned on on the two families of ${\rm D}9^a$ and ${\rm D}9^\alpha$ branes. In what follows, we shall therefore lighten our notation, and refer to ther Chan-Paton labels as $p_a$ and $q_a$, respectively.

This $\mathbb{Z}_2\times \mathbb{Z}_2 $ orientifold admits E5 instantons wrapping the whole $T^6$, together with three different types of BPS ${\rm E}1_i$ instantons, each wrapping the $T^2_i$ internal torus. Given the geometry of ${\rm O}5$ planes, {\it i.e.} two ${\rm O}5_-$ planes each wrapping $T^2_1$ and $T^2_2$, and one ${\rm O}5_+$ plane wrapping the $T^2_3$, the Chan-Paton labels $r_1$ and $r_2$ for the ${\rm E}1_1$ and ${\rm E}1_2$ are complex, while the Chan-Paton label $r_3$ for the ${\rm E}1_3$
brane is real. The parametrisation of Chan-Paton charges for the Euclidean branes is actually determined by the gauge groups that would live on the ${\rm D}5_i$ branes \cite{aadds}, since they are nothing but regular gauge instantons.

Actually, two different types of ${\rm E}1_{1,2}$ instantons and four types of E1$_3$ instantons exist on this orientifold. However, we shall focus our attention only to those parametrised by
\begin{equation}
\begin{split}
&D_{g;o} = r_1+\bar{r}_1 \,,
\\
&D_{g;g} = i (r_1-\bar{r}_1) \,,
\\
&D_{g;f} = r_1+\bar{r}_1 \,,
\\
&D_{g;h} = -i(r_1-\bar{r}_1) \,,
\\
\end{split}
\qquad
\begin{split}
&D_{f;o} = r_2+\bar{r}_2 \,,
\\
&D_{f;g} = r_2+\bar{r}_2 \,,
\\
&D_{f;f} = i (r_2-\bar{r}_2) \,,
\\
&D_{f;h} = i (r_2-\bar{r}_2) \,,
\\
\end{split}
\end{equation}
and by
\begin{equation}
D_{h;o} = D_{h;g} = D_{h;f} = D_{h;h} = r_3 \,.
\end{equation}
In the following, we write, case-by-case, the relevant annulus and M\"obius strip aplitudes, together with the associated zero modes. In all the amplitudes, the index $a$ label the corresponding E-brane, while the index $b$ refers  to the magnetised D9 branes present in the vacuum configuration. Since on the Euclidean branes we do not turn on any magnetic field, in this context $z^{ab}_i = -z^b_i$.

\subsection{${\rm E}1_1$ instantons}

We consider here ${\rm E}1_1$ instantons with wrapping numbers
\begin{equation}
(m,n)=\{ (0,1)\,,\, (1,0)\,,\, (-1,0)\}\,.
\end{equation}
The spectrum of open-strings stretched between two ${\rm E}1_1$ branes is encoded in the amplitudes
\begin{equation}
\begin{split}
\mathcal{A}_{{\rm E}1_1-{\rm E}1_1}
&=
\frac{1}{8}\int_0^{\infty}\frac{dt}{t}\frac{1}{\eta^2}\Bigg\lbrace
D_{g;o}^2\, T_{oo} \, P_1W_2W_3
\\
&+\left[D_{g;g}^2\, T_{og}\, P_1
+D_{g;f}^2\, T_{of}\, W_2+
D_{g;h}^2\, T_{oh}\, W_3\right] \left(\frac{2\eta}{\theta_2}\right)^2\Bigg\rbrace \,,
\\
\end{split}
\end{equation}
and
\begin{equation}
\begin{split}
\mathcal{M}_{{\rm E}1_1}
&
=
\frac{1}{8}D_{g;o}\int_0^{\infty}\frac{dt}{t}\eta^2\left(\frac{2\eta}{\theta_2}\right)^2\Bigg\lbrace
\hat{T}_{oo}^*\, P_1W_2W_3
\\
&
+\left[ -\hat{T}_{og}^*\, P_1 +\hat{T}_{of}^* \, W_2-
\hat{T}_{oh}^*\, W_3\right]\left(\frac{2\eta}{\theta_2}\right)^2\Bigg\rbrace
\,.
\\
\end{split}
\end{equation}
Using the definition and the expansion of the $T_{\mu\nu}$ amplitudes given in Appendix A, we can derive the structure of the neutral zero modes living on the ${\rm E}1_1$ branes
\begin{equation}
\begin{split}
\mathcal{A}_{{\rm E}1_1-{\rm E}1_1}^{(0)}+\mathcal{M}_{{\rm E}1_1}^{(0)}
&=
\frac{1}{2} \left[ r_1(r_1+1) + \bar{r}_1(\bar{r}_1+1) \, \right] \left( O_4O_2V_2O_2  - S_4C_2S_2C_2\right)
\\
&-\frac{1}{2} \left[r_1(r_1-1)+\bar{r}_1(\bar{r}_1-1)\right]\, C_4S_2C_2S_2
\\
&+r_1\bar{r}_1\left[V_4O_2O_2O_2 -(S_4S_2S_2S_2+C_4C_2C_2C_2)\right] \,.
\\
\end{split}
\end{equation}
Therefore, a single ${\rm E}1_1$ instanton has four neutral fermionic zero modes, since the corresponding CP ``group'' is $U(1)$.
As a result, it cannot generate any non-perturbative correction to the superpotential, unless suitable interactions and/or fluxes are turned on \cite{petersson}, \cite{inaki} \cite{Billo':2008sp}.

Open strings stretched between ${\rm E}1_1$ and magnetised D9 branes, are encoded in the annulus amplitude
\begin{equation}
\begin{split}
\mathcal{A}_{{\rm E}1_1-{\rm D}9}
&=\frac{1}{4}\int_0^{\infty}\frac{dt}{t}\eta^2\left(\frac{\eta}{\theta_4}\right)^2
\\
&\times \Bigg\lbrace
I^{ab}\left[(r_1+{\bar r}_1)
 ( \bar{p}_b+\bar{q}_b) T^{57}_{oo} (z_i^{ab}\tau) + {\rm c.c.} \right]
\frac{i\eta}{\theta_1(z_1^{ab}\tau)}\prod_{i=2,3}\frac{\eta}{\theta_4(z_i^{ab}\tau)}
\\
&+S_g^{ab}I_1^{ab}\left[(r_1-{\bar
    r}_1)(\bar{p}_b+\bar{q}_b)T_{og}^{57} (z_i^{ab}\tau)
 - {\rm c.c.} \right]
\frac{i\eta}{\theta_1(z_1^{ab}\tau)}\frac{\eta}{\theta_3(z_2^{ab}\tau)}\frac{\eta}{\theta_3(z_3^{ab}\tau)}
\\
&+S_f^{ab}I_2^{ab}\left[-(r_1+{\bar
r}_1)(\bar{p}_b-\bar{q}_b)iT_{of}^{57} (z_i^{ab}\tau)
 + {\rm c.c.} \right]
\frac{\eta}{\theta_2(z_1^{ab}\tau)}\frac{\eta}{\theta_4(z_2^{ab}\tau)}\frac{\eta}{\theta_3(z_3^{ab}\tau)}
\\
&+S_h^{ab}I_3^{ab}\left[-(r_1-{\bar
    r}_1)(\bar{p}_b-\bar{q}_b)iT_{oh}^{57} (z_i^{ab}\tau) - {\rm
    c.c.} \right]
\frac{\eta}{\theta_2(z_1^{ab}\tau)}\frac{\eta}{\theta_3(z_2^{ab}\tau)}\frac{\eta}{\theta_4(z_3^{ab}\tau)}
\Bigg\rbrace \,, \\
\end{split}
\end{equation}
from which we derive the following charged zero modes
\begin{equation}
\begin{split}
\mathcal{A}_{{\rm E}1_1-{\rm D}9}^{(0)} &= r_1\bar
p_b\left(-O_4S_2O_2O_2\right)\frac{1}{4}\left(I^{ab}+S_g^{ab}I_1^{ab}+S_f^{ab}I_2^{ab}+S_h^{ab}I_3^{ab}\right)
\\
&+r_1\bar
q_b\left(-O_4S_2O_2O_2\right)\frac{1}{4}\left(I^{ab}+S_g^{ab}I_1^{ab}-S_f^{ab}I_2^{ab}-S_h^{ab}I_3^{ab}\right)
\\
&+\bar r_1\bar
p_b\left(-O_4S_2O_2O_2\right)\frac{1}{4}\left(I^{ab}-S_g^{ab}I_1^{ab}+S_f^{ab}I_2^{ab}-S_h^{ab}I_3^{ab}\right)
\\
&+\bar r_1\bar
q_b\left(-O_4S_2O_2O_2\right)\frac{1}{4}\left(I^{ab}-S_g^{ab}I_1^{ab}-S_f^{ab}I_2^{ab}+S_h^{ab}I_3^{ab}\right)
\,.
\\
\end{split}
\end{equation}


\subsection{${\rm E}1_2$ instantons}

We consider here ${\rm E}1_2$ instantons with wrapping numbers
\begin{equation}
(m,n)=\{ (-1,0) \,,\, (0,1)\,,\, (1,0) \}\,.
\end{equation}
The spectrum of open-strings stretched between two ${\rm E}1_2$ branes is encoded in the amplitudes
\begin{equation}
\begin{split}
\mathcal{A}_{{\rm E}1_2-{\rm E}1_2}
&=
\frac{1}{8}\int_0^{\infty}\frac{dt}{t}\frac{1}{\eta^2}\Bigg\lbrace
D_{f;o}^2\, T_{oo}\, W_1P_2W_3
\\
&+\left[ D_{f;g}^2\, T_{og}\, W_1 +D_{f;f}^2\, T_{of}\, P_2+
D_{f;h}^2\, T_{oh}\, W_3\right] \left(\frac{2\eta}{\theta_2}\right)^2\Bigg\rbrace
\\
\end{split}
\end{equation}
and
\begin{equation}
\begin{split}
\mathcal{M}_{{\rm E}1_2}
&=
\frac{1}{8}D_{f;o}\int_0^{\infty}\frac{dt}{t}\eta^2\left(\frac{2\eta}{\theta_2}\right)^2\Bigg\lbrace
\hat{T}_{oo}^* \, W_1P_2W_3
\\
&+\left[ \hat{T}_{og}^*\, W_1 -\hat{T}_{of}^*\, P_2-
\hat{T}_{oh}^*\, W_3\right] \left(\frac{2\eta}{\theta_2}\right)^2\Bigg\rbrace
\,.
\\
\end{split}
\end{equation}
Using the definition and the expansion of the $T_{\mu\nu}$ amplitudes given in Appendix A, we can derive the structure of the neutral zero modes living on the ${\rm E}1_2$ branes
\begin{equation}
\begin{split}
\mathcal{A}_{{\rm E}1_2-{\rm E}1_2}^{(0)}+\mathcal{M}_{{\rm E}1_2}^{(0)}
&= \frac{1}{2}
\left[ r_2(r_2+1)+\bar{r}_2(\bar{r}_2+1)\right] \left( O_4V_2O_2O_2 - S_4S_2C_2C_2 \right)
\\
&- \frac{1}{2} \left[ r_2(r_2-1) + \bar{r}_2(\bar{r}_2-1) \right]\, C_4C_2S_2S_2
\\
&+r_2\bar{r}_2\left[V_4O_2O_2O_2 -(S_4S_2S_2S_2+C_4C_2C_2C_2)\right] \,.
\\
\end{split}
\end{equation}
Therefore, a single ${\rm E}1_2$ instanton has four neutral fermionic zero modes, since the corresponding CP ``group'' is $U(1)$.
As a result, it cannot generate any non-perturbative correction to the superpotential, unless suitable interactions and/or fluxes are turned on \cite{petersson}, \cite{inaki}, \cite{Billo':2008sp}.

Charged zero modes, corresponding to open strings stretched between ${\rm E}1_2$ and magnetised D9 branes, are encoded in the annulus amplitude
\begin{equation}
\begin{split}
\mathcal{A}_{{\rm E}1_2-{\rm D}9}
&=\frac{1}{4}\int_0^{\infty}\frac{dt}{t}\eta^2\left(\frac{\eta}{\theta_4}\right)^2
\\
&\times \Bigg\lbrace
I^{ab}\left[(r_2+{\bar
    r}_2)(\bar{p}_b+\bar{q}_b) T_{oo}^{58}(z_i^{ab}\tau) + {\rm c.c.} \right]
\frac{i\eta}{\theta_1(z_2^{ab}\tau)}\prod_{i=1,3}\frac{\eta}{\theta_4(z_i^{ab}\tau)}
\\
&+S_g^{ab}I_1^{ab}\left[-(r_2+{\bar r}_2)
  (\bar{p}_b+\bar{q}_b)iT_{og}^{58}(z_i^{ab}\tau)
 + {\rm c.c.} \right]
\frac{\eta}{\theta_4(z_1^{ab}\tau)}\frac{\eta}{\theta_2(z_2^{ab}\tau)}\frac{\eta}{\theta_3(z_3^{ab}\tau)}
\\
&+S_f^{ab}I_2^{ab}\left[ (r_2 - {\bar r}_2)
  (\bar{p}_b-\bar{q}_b)T_{of}^{58}(z_i^{ab}\tau)
- {\rm c.c.} \right]
\frac{\eta}{\theta_3(z_1^{ab}\tau)}\frac{i\eta}{\theta_1(z_2^{ab}\tau)}\frac{\eta}{\theta_3(z_3^{ab}\tau)}
\\
&+S_h^{ab}I_3^{ab}\left[(r_2-{\bar r}_2)
  (\bar{p}_b-\bar{q}_b)iT_{oh}^{58}(z_i^{ab}\tau)
+ {\rm c.c.} \right]
\frac{\eta}{\theta_3(z_1^{ab}\tau)}\frac{\eta}{\theta_2(z_2^{ab}\tau)}\frac{\eta}{\theta_4(z_3^{ab}\tau)}
\Bigg\rbrace \,,
\\
\end{split}
\end{equation}
from which we derive the following charged zero modes
\begin{equation}
\begin{split}
\mathcal{A}_{{\rm E}1_2-{\rm D}9}^{(0)} &= r_2\bar p_b\left(-O_4 O_2
S_2
O_2\right)\frac{1}{4}\left(I^{ab}+S_g^{ab}I_1^{ab}+S_f^{ab}I_2^{ab}+S_h^{ab}I_3^{ab}\right)
\\
&+r_2\bar q_b\left(-O_4 O_2
  S_2O_2\right)\frac{1}{4}\left(I^{ab}+S_g^{ab}I_1^{ab}-S_f^{ab}I_2^{ab}-S_h^{ab}I_3^{ab}\right)
\\
&+\bar r_2\bar p_b\left(-O_4 O_2 S_2
O_2\right)\frac{1}{4}\left(I^{ab}+S_g^{ab}I_1^{ab}-S_f^{ab}I_2^{ab}-S_h^{ab}I_3^{ab}\right)
\\
&+\bar r_2\bar q_b\left(-O_4 O_2
S_2O_2\right)\frac{1}{4}\left(I^{ab}+S_g^{ab}I_1^{ab}+S_f^{ab}I_2^{ab}+S_h^{ab}I_3^{ab}\right)
 \,.
 \\
\end{split}
\end{equation}

\subsection{${\rm E}1_3$ instantons}

We consider here ${\rm E}1_3$ instantons with wrapping numbers
\begin{equation}
(m,n)=\{ (1,0) \,,\, (-1,0)\,,\, (0,1) \}\,.
\end{equation}
The spectrum of open-strings stretched between two ${\rm E}1_3$ branes is encoded in the amplitudes
\begin{equation}
\begin{split}
\mathcal{A}_{{\rm E}1_3-{\rm E}1_3}
&=
\frac{1}{8}\int_0^{\infty}\frac{dt}{t}\frac{1}{\eta^2}\Bigg\lbrace
D_{h;o}^2\, T_{oo}\, W_1W_2P_3
\\
&+\left[ D_{h;g}^2\, T_{og}\, W_1 +D_{h;f}^2\, T_{of}\, W_2+
D_{h;h}^2\, T_{oh}\, P_3\right] \left(\frac{2\eta}{\theta_2}\right)^2\Bigg\rbrace \,,
\\
\end{split}
\end{equation}
and
\begin{equation}
\begin{split}
\mathcal{M}_{{\rm E}1_3}
& =
\frac{1}{8} D_{h;o}\int_0^{\infty}\frac{dt}{t}\eta^2\left(\frac{2\eta}{\theta_2}\right)^2\Bigg\lbrace
-\hat{T}_{oo}^*\, W_1W_2P_3
\\
& +\left[-\hat{T}_{og}^* \, W_1 -\hat{T}_{of}^*\, W_2-
\hat{T}_{oh}^*\, P_3\right]
\left(\frac{2\eta}{\theta_2}\right)^2\Bigg\rbrace \,.
\\
\end{split}
\end{equation}
Using the definition and the expansion of the $T_{\mu\nu}$ amplitudes given in Appendix A, we can derive the structure of the neutral zero modes living on the ${\rm E}1_3$ branes
\begin{equation}
\begin{split}
\mathcal{A}_{{\rm E}1_3-{\rm E}1_3}^{(0)}+\mathcal{M}_{{\rm E}1_3}^{(0)}
& = \frac{r_3(r_3+1)}{2} \left( V_4O_2O_2O_2 - C_4C_2C_2C_2 \right)
\\
&-\frac{r_3(r_3-1)}{2}\, S_4S_2S_2S_2
\,.
\\
\end{split}
\end{equation}
Here we see the crucial difference with respect to the ${\rm E}1_{1,2}$ instantons. A single ($r_3=1$) $SO(1)$ ${\rm E}1_3$ instanton  has only two neutral zero modes.  Therefore,  it can generate by itself non-perturbative, single-instanton, contributions to the superpotential.

Open strings stretched between ${\rm E}1_3$ and magnetised D9 branes, are encoded in the annulus amplitude
\begin{equation}
\begin{split}
\mathcal{A}_{{\rm E}1_3-{\rm D}9}
&=\frac{1}{4}\int_0^{\infty}\frac{dt}{t}\eta^2\left(\frac{\eta}{\theta_4}\right)^2\Bigg\lbrace
I^{ab}\left[r_3(\bar{p}_b+\bar{q}_b) T_{oo}^{59}(z_i^{ab}\tau)+
{\rm c.c.} \right]
\prod_{i=1,2}\frac{\eta}{\theta_4(z_i^{ab}\tau)}\frac{i\eta}{\theta_1(z_3^{ab}\tau)}
\\
&+S_g^{ab}I_1^{ab}\left[-r_3(\bar{p}_b+\bar{q}_b)iT_{og}^{59}(z_i^{ab}\tau)
  + {\rm c.c.} \right]
\frac{\eta}{\theta_4(z_1^{ab}\tau)}\frac{\eta}{\theta_3(z_2^{ab}\tau)}\frac{\eta}{\theta_2(z_3^{ab}\tau)}
\\
&+S_f^{ab}I_2^{ab}\left[-r_3(\bar{p}_b-\bar{q}_b)iT_{of}^{59}(z_i^{ab}\tau)
+ {\rm c.c.} \right]
\frac{\eta}{\theta_3(z_1^{ab}\tau)}\frac{\eta}{\theta_4(z_2^{ab}\tau)}\frac{\eta}{\theta_2(z_3^{ab}\tau)}
\\
&+S_h^{ab}I_3^{ab}\left[r_3(\bar{p}_b-\bar{q}_b)T_{oh}^{59}(z_i^{ab}\tau)
  + {\rm c.c.} \right] \frac{\eta}{\theta_3(z_1^{ab}\tau)}\frac{\eta}{\theta_3(z_2^{ab}\tau)}
\frac{i\eta}{\theta_1(z_3^{ab}\tau)}\Bigg\rbrace
\,. \\
\end{split}
\end{equation}
The anti-commuting charged zero modes, corresponding to $O_4 O_2 O_2 S_2$, are displayed in table \ref{Multinst}, and are responsible for the generation of non-perturbative contributions to the superpotential.

\subsection{${\rm E}5$ instantons}\label{e5}

E5 instantons can in principle be endowed with a non trivial magnetised background compatible with the supersymmetry condition
(\ref{SUSY}). Supersymmetric fluxes are thus chosen so that $z_1^a,\ z_2^a>0$ and $z_3^a<0$. In terms of the following parametrisation of the Chan-Paton charges\footnote{Since they
  are gauge instantons for D9 branes, there are four different E5
  instantons. We display for illustration, as before, only one of them.}
\begin{equation}
M_{a,o}=r_a+\bar{r}_a \, , \qquad M_{a,g}=i(r_a-\bar{r}_a) \,,
\qquad M_{a,f}=i(r_a-\bar{r}_a) \, , \qquad M_{a,h}=r_a+\bar{r}_a
\,,
\end{equation}
and of the $T_{\mu\nu}$ defined in Appendix A, the relevant one-loop amplitudes are as follows.

For open string stretched between two ${\rm E}5$ branes, one finds the following annulus
\begin{equation}
\begin{split}
&\mathcal{A}_{{\rm E}5^{(a)}-{\rm E}5^{(a)}}
=\frac{1}{4}\int_0^{\infty}\frac{dt}{t}\Bigg\lbrace  r_a\bar r_a \Bigg[\tilde
P_1\tilde P_2\tilde P_3T_{oo}(0)
\\
&\qquad+\left(\tilde
P_1T_{og}(0)
+\tilde P_2T_{of}(0)+\tilde P_3T_{oh}(0)\right)\left(\frac{2\eta}{\theta_2}\right)^2\Bigg]
\\
&\qquad+I^{aa'}\left[\frac{r_a^2}{2}T_{oo}(2z_i^a\tau)+\frac{\bar
r_a^2}{2}T_{oo}(-2z_i^a\tau)\right]\prod_{i=1}^3\frac{i\eta}{\theta_1(2z_i^a\tau)}
\\
&\qquad-4I_1^{aa'}\left[\frac{r_a^2}{2}T_{og}(2z_i^a\tau)+\frac{\bar
r_a^2}{2}T_{og}(-2z_i^a\tau)\right]\frac{i\eta}{\theta_1(2z_1^a\tau)}\prod_{i=2,3}\frac{\eta}{\theta_2(2z_i^a\tau)}
\\
&\qquad-4I_2^{aa'}\left[\frac{r_a^2}{2}T_{of}(2z_i^a\tau)+\frac{\bar
r_a^2}{2}T_{of}(-2z_i^a\tau)\right]\frac{i\eta}{\theta_1(2z_2^a\tau)}\prod_{i=1,3}\frac{\eta}{\theta_2(2z_i^a\tau)}
\\
&\qquad+4I_3^{aa'}\left[\frac{r_a^2}{2}T_{oh}(2z_i^a\tau)+\frac{\bar
r_a^2}{2}T_{oh}(-2z_i^a\tau)\right]\frac{i\eta}{\theta_1(2z_3^a\tau)}\prod_{i=1,2}\frac{\eta}{\theta_2(2z_i^a\tau)}\Bigg\rbrace\frac{1}{\eta^2} \,,
\\
\end{split}
\end{equation}
and M\"obius amplitude
\begin{equation}
\begin{split}
\mathcal{M}_{{\rm E}5^{(a)}}
&=-\int_0^{\infty}\frac{dt}{t}\hat{\eta}^2\left(\frac{2\hat{\eta}}{\hat{\theta}_2}\right)^2\Bigg\lbrace
\prod_{i=1}^3(m^{(a)}_i)\left[r_a\,\hat{T}_{oo}^*(2z_i^a\tau)+\bar{r}_a\,\hat{T}_{oo}^*(-2z_i^a\tau)\right]\prod_{i=1}^3
\frac{i\hat{\eta}}{\hat{\theta}_1(2z_i^a\tau)}
\\
&-\epsilon_1\left[r_a\, \hat{T}_{og}^*(2z_i^a\tau)+\bar{r}_a\, \hat{T}_{og}^*(-2z_i^a\tau)\right]\frac{im^{(a)}_1\hat{\eta}}{\hat{\theta}_1(2z^a_1\tau)}
\prod_{i=2,3}\frac{n^{(a)}_i\hat{\eta}}{\hat{\theta}_2(2z_i^a\tau)}
\\
&-\epsilon_2\left[r_a\, \hat{T}_{of}^*(2z_i^a\tau)+\bar{r}_a\, \hat{T}_{of}^*(-2z_i^a\tau)\right]\frac{im^{(a)}_2\hat{\eta}}{\hat{\theta}_1(2z^a_2\tau)}
\prod_{i=1,3}\frac{n^{(a)}_i\hat{\eta}}{\hat{\theta}_2(2z_i^a\tau)}
\\
&-\epsilon_3\left[r_a\, \hat{T}_{oh}^*(2z_i^a\tau)+\bar{r}_a\, \hat{T}_{oh}^*(-2z_i^a\tau)\right]\frac{im^{(a)}_3\hat{\eta}}{\hat{\theta}_1(2z^a_3\tau)}
\prod_{i=1,2}\frac{n^{(a)}_i\hat{\eta}}{\hat{\theta}_2(2z_i^a\tau)}\Bigg\rbrace
\,.
\\
\end{split}
\end{equation}
We do not write here interactions between {\it different} instantons.  The massless neutral zero modes are then given by
\begin{equation}
\begin{split}
\mathcal{A}_{{\rm E}5^{(a)}-{\rm E}5^{(a)}}^{(0)}+\mathcal{M}_{{\rm E}5^{(a)}}^{(0)}
&= r_a\bar{r_a}\, \left[V_4O_2O_2O_2
-(S_4S_2S_2S_2+C_4C_2C_2C_2)\right]
\\
&+\frac{r_a(r_a+1)}{2}\Big[-O_4O_2O_2V_2\, \frac{1}{8}(I^{aa'}-I^{aO}-4I_1^{aa'}-4I_2^{aa'}+4I_3^{aa'})
\\
&+C_4S_2S_2C_2\, \frac{1}{8}(I^{aa'}+I^{aO}-4I_1^{aa'}-4I_2^{aa'}+4I_3^{aa'})\Big]
\\
&+\frac{\bar{r}_a(\bar{r}_a+1)}{2}\Big[-O_4O_2O_2V_2\, \frac{1}{8}(I^{aa'}-I^{aO}-4I_1^{aa'}-4I_2^{aa'}+4I_3^{aa'})
\\
&+S_4C_2C_2S_2\, \frac{1}{8}(I^{aa'}-I^{aO}-4I_1^{aa'}-4I_2^{aa'}+4I_3^{aa'})\Big]
\\
&+\frac{r_a(r_a-1)}{2}\Big[-O_4O_2O_2V_2\, \frac{1}{8}(I^{aa'}+I^{aO}-4I_1^{aa'}-4I_2^{aa'}+4I_3^{aa'})
\\
&+C_4S_2S_2C_2\, \frac{1}{8}(I^{aa'}-I^{aO}-4I_1^{aa'}-4I_2^{aa'}+4I_3^{aa'})\Big]
\\
&+\frac{\bar{r}_a(\bar{r}_a+1)}{2}\Big[-O_4O_2O_2V_2\, \frac{1}{8}(I^{aa'}+I^{aO}-4I_1^{aa'}-4I_2^{aa'}+4I_3^{aa'})
\\
&+S_4C_2C_2S_2\, \frac{1}{8}(I^{aa'}+I^{aO}-4I_1^{aa'}-4I_2^{aa'}+4I_3^{aa'})\Big]\,.
\\
\end{split}
\end{equation}
Finally, for open strings stretched between ${\rm E}5$ and ${\rm D}9$ branes one finds the annulus amplitude
\begin{equation}
\begin{split}
&\mathcal{A}_{{\rm E}5^{(a)}-{\rm D}9^{(b)}}
=\frac{1}{4}\int_0^{\infty}\frac{dt}{t}\eta^2\left(\frac{\eta}{\theta_4}\right)^2
\\
&\quad\times\Bigg\lbrace
I^{ab}\left[r_a(\bar{p}_b+\bar{q}_b) {T}_{oo}^{56} (z_i^{ab}\tau)+\bar{r}_a(p_b+q_b) {T}^{56} _{oo}(-z_i^{ab}\tau)\right]
\prod_{i=1}^3\frac{i\eta}{\theta_1(z_i^{ab}\tau)}
\\
&+I^{ab'}\left[r_a(p_b+q_b) {T}^{56} _{oo}(z_i^{ab'}\tau)+\bar{r}_a(\bar{p}_b+\bar{q}_b){T}^{56} _{oo}(-z_i^{ab'}\tau)\right]
\prod_{i=1}^3\frac{i\eta}{\theta_1(z_i^{ab'}\tau)}
\\
&+S_g^{ab} I_1^{ab}\left[r_a(\bar{p}_b+\bar{q}_b){T}^{56} _{og}(z_i^{ab}\tau)+\bar{r}_a(p_b+q_b){T}^{56} _{og}(-z_i^{ab}\tau)\right]
\frac{i\eta}{\theta_1(z_1^{ab}\tau)}\prod_{i=2,3}\frac{\eta}{\theta_2(z_i^{ab}\tau)}
\\
&-S_g^{ab} I_1^{ab'}\left[r_a(p_b+q_b){T}^{56} _{og}(z_i^{ab'}\tau)+\bar{r}_a(\bar{p}_b+\bar{q}_b){T}^{56} _{og}(-z_i^{ab'}\tau)\right] \frac{i\eta}{\theta_1(z_1^{ab'}\tau)}\prod_{i=2,3}\frac{\eta}{\theta_2(z_i^{ab'}\tau)}
\\
&+S_f^{ab} I_2^{ab}\left[r_a(\bar{p}_b-\bar{q}_b){T}^{56} _{of}(z_i^{ab}\tau)+\bar{r}_a(p_b-q_b){T}^{56} _{of}(-z_i^{ab}\tau)\right]
\frac{i\eta}{\theta_1(z_2^{ab}\tau)}\prod_{i=1,3}\frac{\eta}{\theta_2(z_i^{ab}\tau)}
\\
&-S_f^{ab} I_2^{ab'}\left[r_a(p_b-q_b){T}^{56} _{of}(z_i^{ab'}\tau)+\bar{r}_a(\bar{p}_b-\bar{q}_b) {T}^{56} _{of}(-z_i^{ab'}\tau)\right]
\frac{i\eta}{\theta_1(z_2^{ab'}\tau)}\prod_{i=1,3}\frac{\eta}{\theta_2(z_i^{ab'}\tau)}
\\
&+S_h^{ab} I_3^{ab}\left[r_a(\bar{p}_b-\bar{q}_b) {T}^{56} _{oh}(z_i^{ab}\tau)+\bar{r}_a(p_b-q_b) {T}^{56} _{oh}(-z_i^{ab}\tau)\right]
\frac{i\eta}{\theta_1(z_3^{ab}\tau)}\prod_{i=1,2}\frac{\eta}{\theta_2(z_i^{ab}\tau)}
\\
&+S_h^{ab} I_3^{ab'}\left[r_a(p_b-q_b) {T}^{56} _{oh}(z_i^{ab'}\tau)+\bar{r}_a(\bar{p}_b-\bar{q}_b) {T}^{56} _{oh}(-z_i^{ab'}\tau)\right] \frac{i\eta}{\theta_1(z_3^{ab'}\tau)}\prod_{i=1,2}\frac{\eta}{\theta_2(z_i^{ab'}\tau)}\Bigg\rbrace\,.
\\
\end{split}
\end{equation}
If fluxes are chosen such that $z_1^{ab},z_2^{ab}>0$, $z_3^{ab}<0$
and $z_1^{ab'},z_2^{ab'}>0$, $z_3^{ab'}<0$, the massless spectrum is encoded in
\begin{equation}
\begin{split}
\mathcal{A}_{{\rm E}5^{(a)}-{\rm D}9^{(b)}}^{(0)}
&=\bar{r}_ap_b\,
\frac{1}{4} \left(I^{ab}+S_g^{ab}I_1^{ab}+S_f^{ab}I_2^{ab}+S_h^{ab}I_3^{ab}\right)
\,  O_4S_2S_2C_2
\\
&+\bar{r}_aq_b\,
\frac{1}{4} \left(I^{ab}+S_g^{ab}I_1^{ab}-S_f^{ab}I_2^{ab}-S_h^{ab}I_3^{ab}\right)\, O_4S_2S_2C_2
\\
&+\bar{r}_a\bar{p}_b\,
\frac{1}{4} \left(I^{ab'}-S_g^{ab}I_1^{ab'}-S_f^{ab}I_2^{ab'}+S_h^{ab}I_3^{ab'}\right)\, O_4S_2S_2C_2
\\
&+\bar{r}_a\bar{q}_b\,
\frac{1}{4} \left(I^{ab'}-S_g^{ab}I_1^{ab'}+S_f^{ab}I_2^{ab'}-S_h^{ab}I_3^{ab'}\right) \, O_4S_2S_2C_2
\,.
\\
\end{split}
\end{equation}
E$5^{(a)}$ instantons have unitary CP factors and therefore a minimum number
of four zero modes, and are expected to play a role in
generating non-perturbative effects on the D9-brane gauge theories.

\subsection{Partition functions involving non-magnetised D-branes with Wilson lines}\label{par}

In this last Appendix, we summarise the one loop amplitudes needed to extract the spectrum for the $U(2)^2\times U(2)^2\times USp(4)^2\times USp(4)^2$ model studied in Section 4.3.

The annulus and M\"obius amplitudes for open strings ending on pairs of non-magnetised D9 branes (${\rm D}9_{\rm nm}$ for short) with Wilson line $a$ in the third torus, and on pairs of ${\rm D}5_1$ branes shifted away from the origin of the second torus by a distance proportional to $a'$, are given by
\begin{equation}
\begin{split}
\mathcal{A}
&=\frac{1}{8}\int_{0}^{\infty}\frac{dt}{t^3}\frac{1}{\eta^2}\Bigg\lbrace\Bigg[
\frac{N^2}{2} \left(P_3
+\frac{1}{2}P_{m_3+2a}+\frac{1}{2}P_{m_3-2a}\right)P_1P_2
\\
&+\frac{D_1^2}{2} \left(W_2+\frac{1}{2}W_{n_2+2a'}+\frac{1}{2}W_{n_2-2a'}\right)P_1W_3\Bigg]T_{oo}
\\
&+2ND_1P_1T_{go}\left(\frac{\eta}{\theta_4}\right)^2
\\
&+\frac{D_{1f}^2}{2}\left(W_2+\frac{W_{n_2+2a'}+W_{n_2-2a'}}{2}\right)
T_{of}\left(\frac{2\eta}{\theta_2}\right)^2
\\
&+\frac{N_{h}^2}{2}\left(P_3+\frac{P_{m_3+2a}+P_{m_3-2a}}{2}\right)
T_{oh}\left(\frac{2\eta}{\theta_2}\right)^2 \Bigg\rbrace
\,,
\\
\end{split}
\end{equation}
and
\begin{equation}
\begin{split}
\mathcal{M}
&=-\frac{1}{8}\int_{0}^{\infty}\frac{dt}{t^3}\frac{1}{\eta^2}\Bigg\lbrace\Bigg[\frac{N}{2}P_1P_2(P_{m_3+2a}+P_{m_3-2a})
\\
&+\frac{D_1}{2}P_1(W_{n_2+2a'}+W_{n_2-2a'})W_3\Bigg]T_{oo}
\\
&+\frac{D_1}{2}(W_{n_2+2a'}+W_{n_2-2a'})T_{of}\left(\frac{2\eta}{\theta_2}\right)^2
+\frac{N}{2}(P_{m_3+2a}+P_{m_3-2a})T_{oh}\left(\frac{2\eta}{\theta_2}\right)^2
\\
&-(N+D_1)P_1T_{og}\left(\frac{2\eta}{\theta_2}\right)^2
-(NP_2T_{of}+D_1W_3T_{oh})\left(\frac{2\eta}{\theta_2}\right)^2\Bigg\rbrace
\,.
\\
\end{split}
\end{equation}
Because of the presence of Wilson-lines, the Chan-Paton factors of the ${\rm D}9_{\rm nm}$ and ${\rm D}5_1$ branes need to be properly rescaled
\begin{equation}
D_1=2(d_1+d_2) \, , \qquad D_{1f}=2(d_1-d_2) \, , \qquad
N=2(n_1+n_2) \, , \qquad N_h=2(n_1-n_2) \,.
\end{equation}
Actually, in a geometric language, this corresponds to  the fact that ${\rm D}9_{\rm nm}$ and the D5$_1$
must appear in suitable multiplets that are exchanged by the orbifold operations.

One also needs to consider open strings stretched between magnetised D9 branes and ${\rm D}9_{\rm nm}$ and $D5_1$ branes. Their spectra are encoded in
\begin{equation}
\begin{split}
\mathcal{A}_{{\rm D}9-{\rm D}9_{\rm nm}}
&= \frac{1}{2}\int_0^{\infty}\frac{dt}{t^3}\frac{1}{\eta^2}\Bigg\lbrace
I^{ab}\left[(p_a+q_a)(n_1+n_2)T_{oo}(z_i^{ab}\tau) + {\rm c.c.} \right]
\prod_{i=1}^3\frac{i\eta}{\theta_1(z_i^{ab}\tau)}
\\
&+I_3^{ab}\left[(p_a-q_a)(n_1-n_2)T_{oh}(z_i^{ab}\tau)
+ {\rm c.c.} \right]
\frac{2i\eta}{\theta_1(z_3^{ab}\tau)}\prod_{i=1,2}\frac{\eta}{\theta_2(z_i^{ab}\tau)}\Bigg\rbrace
\,,
\\
\end{split}
\end{equation}
and
\begin{equation}
\begin{split}
\mathcal{A}_{{\rm D}9-{\rm D}5_1}
&=\frac{1}{2}\int_0^{\infty}\frac{dt}{t^3}\frac{1}{\eta^2}\Bigg\lbrace
I^{ab}\left[(p_a+q_a)(d_1+d_2)T_{go}(z_i^{ab}\tau) + {\rm c.c.} \right]
\frac{i\eta}{\theta_1(z_1^{ab}\tau)}\prod_{i=2,3}\frac{\eta}{\theta_4(z_i^{ab}\tau)}\\
&+I_2^{ab}\left[(p_a-q_a)(d_1-d_2)T_{gf}(z_i^{ab}\tau) - {\rm c.c.} \right]
\frac{2\eta^3}{\theta_2(z_1^{ab}\tau)\theta_4(z_2^{ab}\tau)\theta_3(z_3^{ab}\tau)}\Bigg\rbrace
\,, \\
\end{split}
\end{equation}
where the index $b$ refers, universally, to the ${\rm D}9_{\rm nm}$ and the ${\rm D}5_1$ branes.

Turning to the instantonic branes, there are no charged zero-modes associated to open strings stretched between the rigid
instanton  and the ${\rm D}9_{\rm nm}$ and $D5_1$ branes because of the Wilson line $a$ and the shift $a'$. This can be easily
seen from the corresponding annulus amplitudes
\begin{equation}
\begin{split}
\mathcal{A}_{{\rm E}1_3-{\rm D}9_{\rm{nm}}}
&=\frac{1}{4}\int_0^{\infty}\frac{dt}{t}\eta^2\left(\frac{\eta}{\theta_4}\right)^2
\Bigg\lbrace
r_3(n_1+n_2)T_{oo}^{59}\left(\frac{\eta}{\theta_4}\right)^2
\\
&+r_3(n_1-n_2)T_{oh}^{59}\left(\frac{\eta}{\theta_3}\right)^2\Bigg\rbrace
\left(P_{m_3+a}+P_{m_3-a}\right) \,,
\\
\end{split}
\end{equation}
and
\begin{equation}
\begin{split}
\mathcal{A}_{{\rm E}1_3-{\rm D}5_{1}}
&=\frac{1}{4}\int_0^{\infty}\frac{dt}{t}\eta^2\left(\frac{\eta}{\theta_4}\right)^2
\Bigg\lbrace
r_3(d_1+d_2)T_{oo}^{58}\left(\frac{\eta}{\theta_4}\right)^2
\\
&+r_3(d_1-d_2)T_{of}^{58}\left(\frac{\eta}{\theta_3}\right)^2\Bigg\rbrace
\left(W_{n_2+a'}+W_{n_2-a'}\right) \,.
\\
\end{split}
\end{equation}
Notice that in this last amplitude we have used the $T_{\mu \nu}^{58}$ pertaining to open strings stretched between D9 and ${\rm E}1_2$ branes. This is correct since the pairs D9--${\rm E}1_2$ and ${\rm D}5_1$--${\rm E}1_3$ have precisely the same structure of  Neumann and Dirichlet boundary conditions.


\end{document}